\documentclass[preprint,3p,10pt]{elsarticle}
\usepackage{hyperref}

\usepackage[T1]{fontenc}
\usepackage[utf8]{luainputenc}
\usepackage{verbatim}
\usepackage{amsmath}
\usepackage{amssymb}
\usepackage{graphicx}
\usepackage{esint}
\usepackage{times}
\usepackage{color}
\journal{CR Physique}

\begin{document}
\let\vaccent=\v{% rename builtin command \v{} to \vaccent{}}\global\long\def\v#1{\ensuremath{\mathbf{#1}}}
 % for vectors
\global\long\def\gv#1{\ensuremath{\mbox{\boldmath\ensuremath{#1}}}}
 % for vectors of Greek letters
\global\long\def\uv#1{\ensuremath{\mathbf{\hat{#1}}}}
 % for unit vector
\global\long\def\abs#1{\left| #1 \right|}
 % for absolute value
\global\long\def\avg#1{\left< #1 \right>}
 % for average
\let\underdot=\d{% rename builtin command \d{} to \underdot{}}\global\long\def\d#1#2{\frac{d#1}{d#2}}
 % for derivatives
\global\long\def\dd#1#2{\frac{d^{2}#1}{d#2^{2}}}
 % for double derivatives
\global\long\def\pd#1#2{\frac{\partial#1}{\partial#2}}
 % for partial derivatives
\global\long\def\pdd#1#2{\frac{\partial^{2}#1}{\partial#2^{2}}}
 % for double partial derivatives
\global\long\def\pdc#1#2#3{\left( \frac{\partial#1}{\partial#2}\right)_{#3}}
 % for thermodynamic partial derivatives
\global\long\def\op#1{\hat{\mathrm{#1}}}
 % for operators
\global\long\def\ket#1{\left| #1 \right>}
 % for Dirac bras
\global\long\def\bra#1{\left< #1 \right|}
 % for Dirac kets
\global\long\def\braket#1#2{\left< #1 \vphantom{#2}\right| \left. #2 \vphantom{#1}\right>}
 % for Dirac brackets
\global\long\def\matrixel#1#2#3{\left< #1 \vphantom{#2#3}\right| #2 \left| #3 \vphantom{#1#2}\right>}
 % for Dirac matrix elements
\global\long\def\av#1{\left\langle #1 \right\rangle }
 \global\long\def\com#1#2{\left[#1,#2\right]}
 %for commutators
\global\long\def\acom#1#2{\left\{  #1,#2\right\}  }
 %for anti commutators
\global\long\def\grad#1{\gv{\nabla} #1}
 % for gradient
\let\divsymb=\div % rename builtin command \div to \divsymb
\global\long\def\div#1{\gv{\nabla} \cdot#1}
 % for divergence
\global\long\def\curl#1{\gv{\nabla} \times#1}
 % for curl
\let\baraccent=\={% rename builtin command \= to \baraccent}\global\long\def\=#1{\stackrel{#1}{=}}
 % for putting numbers above =
% 
% \title{Strongly interacting photons in arrays of dissipative nonlinear cavities under a frequency-dependent incoherent pumping}
% 
% 
% \author{Jos\'e Lebreuilly}
% \email{jose.lebreuilly@unitn.it}
% \affiliation{D\'epartement de Physique de l'\'Ecole Normale Sup\'erieure, 24 rue Lhomond, 75231 Paris, France}
% \affiliation{INO-CNR BEC Center and Dipartimento di Fisica, Universit\`a di Trento, I-38123 Povo, Italy}
% \author{Michiel Wouters}
% \affiliation{TQC, Universiteit Antwerpen, Universiteitsplein 1, B-2610 Antwerpen, Belgium}
% \author{Iacopo Carusotto}
% \affiliation{INO-CNR BEC Center and Dipartimento di Fisica, Universit\`a di Trento, I-38123 Povo, Italy}

\begin{abstract}
\end{abstract}

\begin{frontmatter}

\title{Towards strongly {correlated} photons in arrays of dissipative nonlinear cavities under a frequency-dependent incoherent pumping}

% %% Group authors per affiliation:
% \address{Radarweg 29, Amsterdam}
% \fntext[myfootnote]{Since 1880.}

\author[paris,tn]{Jos\'e Lebreuilly\corref{mycorrespondingauthor}}
\ead{jose.lebreuilly@unitn.it}

\author[antwerp]{Michiel Wouters}
\author[tn]{Iacopo Carusotto\fnref{postaladdress}}
\ead{carusott@science.unitn.it}
\cortext[mycorrespondingauthor]{Corresponding author}
\fntext[postaladdress]{Tel. +39 0461 283925. Fax. +39 0461 282014 }

\address[paris]{D\'epartement de Physique de l'\'Ecole Normale Sup\'erieure, 24 rue Lhomond, 75231 Paris, France}
\address[tn]{INO-CNR BEC Center and Dipartimento di Fisica, Universit\`a di Trento, via Sommarive 14, I-38123 Povo, Italy}
\address[antwerp]{TQC, Universiteit Antwerpen, Universiteitsplein 1, B-2610 Antwerpen, Belgium}

\begin{abstract}
We report a theoretical study of a quantum optical model consisting of an array of strongly nonlinear cavities incoherently pumped by an ensemble of population-inverted two-level atoms. Projective methods are used to eliminate the atomic dynamics and write a generalized master equation for the photonic degrees of freedom only, where the frequency-dependence of gain introduces non-Markovian features. In the simplest single cavity configuration, this pumping scheme gives novel optical bistability effects and allows for the selective generation of Fock states with a well-defined photon number. For many cavities in a weakly non-Markovian limit, the non-equilibrium steady state recovers a Grand-Canonical statistical ensemble at a temperature determined by the effective atomic linewidth. For a two-cavity system in the strongly nonlinear regime, signatures of a Mott state with one photon per cavity are found.
\end{abstract}


\end{frontmatter}

\section{Introduction}

The study of quantum many-body systems is one of the most active fields of modern condensed-matter physics. Among the most celebrated effects, we can mention frictionless flows in superfluid and superconducting systems and the geometrical quantization features of the fractional quantum Hall effect. While this physics was traditionally studied in liquid Helium samples~\cite{pinesnozieres,leggett2004}, in atomic nuclei~\cite{ringschuck}, in quark-gluon plasmas~\cite{yagi,sinha}, or in electron gases confined in solid-state devices~\cite{Schrieffer1964,Mahan1990,Tinkham2004,Yoshioka2002}, {the last two decades have witnessed impressive advances} using ultra-cold atomic gases trapped in magnetic or optical traps~\cite{Dalfovo1999,Bloch2008,Giorgini2008}. 

In the last few years, a growing community has started investigating many-body effects in the novel context of the so-called quantum fluids of light~\cite{RMP}, i.e. assemblies of many photons confined in suitable optical devices, where effective photon-photon interactions arise from the optical nonlinearity of the medium. After the pioneering studies of Bose-Einstein condensation~\cite{BEC} and superfluidity~\cite{superfluidity} effects in dilute photon gases in weakly nonlinear media, a great interest is presently being devoted to strongly nonlinear systems, where even single photons are able to appreciably affect the optical properties of the system. 

The most celebrated example of such physics is the photon blockade effect~\cite{imamoglu}, where the presence of a single photon in a cavity is able to {detune the cavity} frequency away from the pump laser, so that photons behave as effectively impenetrable particles. Experimental realizations of this idea have been reported by several groups using very different material platforms, from single atoms in macroscopic cavities~\cite{birnbaum}, to single quantum dots in photonic crystal cavities~\cite{vuckovic,Reinhard}, to single Josephson qubits in circuit QED devices for microwaves~\cite{CiQED-blockade,CiQED_rev}.

Scaling up to arrays of many cavities coupled by photon tunneling is presently a hot challenge in experimental physics{, as it would realize a Bose-Hubbard model for photons where the photon blockade effect may lead to a rich} physics, including the superfluid to Mott-insulator phase transition at a commensurate filling or Tonks-Girardeau gases of impenetrable photons in one-dimensional continuum models. The first works on strongly correlated photons were restricted to quasi-equilibrium regimes where the photon loss rate is much slower than the internal dynamics of the gas so that the system has time to thermalize and/or be adiabatically transfered to the desired strongly correlated state~\cite{photonMI,TGHarvard}. While this assumption might be satisfied in suitably designed circuit-QED devices in the microwave domain, radiative losses are hardly negligible in realistic optical cavities in the infrared or visible domain, {so that} thermalization is {generally} far from being granted~\cite{RMP,CiQED_rev}.

As a result, a very active attention has been {recently} devoted to the peculiar non-equilibrium effects that arise for realistic loss rates. Starting from the pioneering work on photon blockade in non-equilibrium photonic Josephson junctions~\cite{gerace-fazio}, the interest has {been focused} on the study of schemes to generate strongly correlated many-body states {in the very non-equilibrium context of photon systems, where the steady-state is not determined by a thermal equilibrium condition, but by a dynamical balance of driving and losses.}

The first such scheme proposed in~\cite{TGnoi} was based on a coherent pumping: provided the different many-body states are sufficiently separated in energy, many-photon processes driven by the coherent external laser are able to selectively address each many-body state as done in optical spectroscopy of atomic levels. In this way, the non-equilibrium condition is no longer just a hindrance, but offers new perspectives, as it allows to individually probe each excited state. Furthermore, the appreciable radiative losses make microscopic information on the many-body wavefunction be directly encoded in the quantum coherence of the secondary emission from the device~\cite{FQH,UWC,Braid}. While this coherent pumping scheme offers a viable way to generate and control few photon states in small arrays, its efficiency is restricted to mesoscopic systems where the different states are well-separated in energy. Moreover, this scheme intrinsically leads to coherent superpositions of states of different photon number: while this feature is intriguing in view of observing many-body braiding phases~\cite{Braid}, it is not ideally suited to generate states with a well-defined photon number such as Mott-insulator states.

The identification of new schemes that do not suffer from these limitations is therefore of great importance in view of experiments. In the present work we study the potential of frequency-dependent gain processes to selectively generate strongly correlated states of photons in arrays of strongly nonlinear cavities. The frequency-dependence of amplification is a well-known fact of laser physics and is often exploited to choose and stabilize a desired lasing mode~\cite{laser}. In the last years, a series of works by our groups~\cite{PRL10,Chiocchetta13} have explored its effect on exciton-polariton Bose-Einstein condensation experiments, in particular questioning the apparent thermalization of the non-condensed fraction~\cite{Chiocchetta14,Bajoni,Weitz,Kirton14,alessio2}. All these works were however restricted to the weakly interacting regime where quantum fluctuations can be treated in the input-output language by means of a Bogoliubov-like linearized theory around the mean-field. Here we tackle the far more difficult case of strong nonlinearities, which requires including the non-Markovian features due to the frequency-dependent gain into the many-body master equation for the strongly interacting photons and then to solve the quantum many-body theory of the generalized driven-dissipative Bose-Hubbard model. 

In the last years, similar questions have been theoretically addressed by several groups. Just to mention a few of them, a scheme to obtain a thermal state at finite temperature with a non-vanishing effective chemical potential for photons has been proposed in~\cite{Adhikari} using a clever parametric system-bath coupling with as special eye to circuit-QED and opto-mechanical systems. A further development in this direction~\cite{Kapit} has considered pumping by two-photon processes in the presence of an auxiliary shadow lattice in a circuit-QED architecture: in spite of the complexity of the proposed set-up, the mechanism underlying the stabilization of many-body states is very similar to our frequency-dependent gain. With respect to these proposals and to the engineered dissipations originally proposed for atoms~\cite{diehl} and then extended to photons~\cite{hafezi,domokos} to organic polaritons \cite{Lambrecht,keeling2}, and circuit QED systems \cite{aron,gourgy,tureci}, our approach has the crucial advantage of being based on a quite commonly observed feature of laser and photonic systems such as a frequency-dependent gain.
Finally, a pioneering discussion of the onset of collective coherence in a related model of a cavity array embedding population-inverted atoms has recently appeared in~\cite{Ruiz}, but little attention was paid to the effect of strong nonlinearities nor to the development of a tractable quantum formalism.

The aim of this article is to introduce the readers to the basic physics of a frequency-dependent incoherent pumping and to first illustrate the consequences of the resulting non-Markovianity in the simplest configurations before attacking more complex many-body effects. With this idea in mind, the structure of the article is the following. In Sec.\ref{sec:Presentation-of-the} we present the physical system and we develop the theoretical model based on a master equation for the cavities coupled to the atoms of the gain medium. The projective method to eliminate the atomic degrees of freedom and write a master equation for the photonic density matrix is sketched in Sec.\ref{sec:Closed-master-equation} along the lines of the general theory of~\cite{Breuer}. First application of the method to a single cavity configuration is discussed in Sec.\ref{sec:onecavity} and specific features of the weak and the strong nonlinearity cases are illustrated, e.g. a novel mechanism for optical bistability and the selective generation of Fock states with a well defined photon number. The richer physics of many cavity arrays is discussed in Sec.\ref{sec:Arrays-of-cavities}: In a Markovian regime, the photonic steady state has the surprisingly trivial form of a Grand-Canonical distribution of infinite temperature, and therefore is fully independent of the many-body photonic Hamiltonian. In a weakly non-Markovian regime, an effective Grand-Canonical distribution of finite temperature is obtained even in the absence of thermalization mechanisms; in a strongly nonlinear and non-Markovian regime, signatures of a Mott insulator state with one photon per cavity are illustrated. Conclusions are finally drawn in Sec.\ref{sec:Conclu}. In the Appendices, we provide the details of the derivation of the photonic master equation using projective methods, on the exact stationary state in the Markovian case, on a perturbative expansion of the coherences in the weakly non-Markovian limit, and on further numerical validation of the purely photonic master equation.

\section{The physical system and the theoretical model}\label{sec:Presentation-of-the}

\subsection{The physical system}

In this work, we consider a driven-dissipative Bose-Hubbard model for photons in an array of $k$ coupled nonlinear cavities of natural frequency $\omega_{cav}$. In units such that $\hbar=1$, the Hamiltonian for the isolated system dynamics has the usual form~\cite{RMP,CiQED_rev,Hartmann_rev}:
\begin{equation}
H_{ph}=\sum_{i=1}^{k}\left[\omega_{cav}a_{i}^{\dagger}a_{i}+\frac{U}{2}a_{i}^{\dagger}a_{i}^{\dagger}a_{i}a_{i}\right]{-}\sum_{\avg{i,j}}\left[ Ja_{i}^{\dagger}a_{j}+hc\right].
\end{equation}
They are arranged in a one-dimensional geometry {and} are coupled via tunneling processes with amplitude $J$. Each cavity is assumed to contain a  Kerr nonlinear medium, which induces effective repulsive interactions between photons in the same cavity with an interaction constant $U$ proportional to the Kerr nonlinearity $\chi^{(3)}$.
Dissipative phenomena due the finite transparency of the mirrors and absorption {by the cavity material} are responsible for a finite lifetime of photons, which naturally decay at a rate $\Gamma_{loss}$. 

As mentioned in the introduction, the key novelty of this work with respect to earlier work consists in the different mechanism that is proposed to compensate for losses and replenish the photon population. Instead of a coherent pumping or a very broad-band {amplifying} laser medium, we consider a configuration where a set of $N_{at}$ two level atoms is present in each cavity. Each atom is strongly pumped at a rate $\Gamma_{pump}$, spontaneously decays to its ground state at a rate $\gamma$ and, most importantly, is coupled to the cavity with a Rabi frequency $\Omega_{R}$: as a result, the atoms provide an incoherent pumping of the cavities, with a frequency-dependent rate centered at the atomic frequency $\omega_{at}$. Our choice of two different physical mechanisms for nonlinearity and pumping (for example, two different atomic species) allows us to to tune independently photonic interactions and emission.

%If we remain in small emission rate, maybe we don't need Na atoms ?

The free evolution of the atoms and their coupling to the cavities are described by the following Hamiltonian terms,
\begin{eqnarray}
H_{at}&=&\sum_{i=1}^{k}\sum_{l=1}^{N_{at}}\omega_{at}\sigma_{i}^{+(l)}\sigma_{i}^{-(l)} \\ 
H_{I}&=&\Omega_{R}\sum_{i=1}^{k}\sum_{l=1}^{N_{at}}\left[a_{i}^{\dagger}\sigma_{i}^{-(l)}+a_{i}\sigma_{i}^{+(l)}\right]:
\end{eqnarray}
the atomic frequency $\omega_{at}$ is assumed to be in the vicinity (but not necessarily resonant) with the cavity mode and the atom-cavity coupling is assumed to be weak enough $\Omega_R\ll \omega_{at},\omega_{cav}$ to be far from the ultra-strong coupling regime~\cite{bastard} and from any superradiant Dicke transition~\cite{dicke}.

As usual, the dissipative dynamics under the effect of the pumping and decay processes can be described in terms of a master equation for the density matrix $\rho$ of the whole atom-cavity system, 
\begin{equation}
\partial_{t}\rho=\frac{1}{i}\com{H_{ph}+H_{at}+H_{I}}{\rho}+\mathcal{L}(\rho),\label{eq:evinitio}
\end{equation}
where the different dissipative processes are summarized in the Lindblad super-operator $\mathcal{L}=\mathcal{L}_{pump}+\mathcal{L}_{loss,\, at}+\mathcal{L}_{loss,\, cav}$, with 
\begin{eqnarray}
\mathcal{L}_{pump}  &=&  \frac{\Gamma_{pump}}{2}\sum_{i=1}^{k}\sum_{l=1}^{N_{at}}\left[2\sigma_{i}^{+(l)}\rho\sigma_{i}^{-(l)}-\sigma_{i}^{-(l)}\sigma_{i}^{+(l)}\rho-\rho\sigma_{i}^{-(l)}\sigma_{i}^{+(l)}\right], \\
\mathcal{L}_{loss,\, at} & = & \frac{\gamma}{2}\sum_{i=1}^{k}\sum_{l=1}^{N_{at}}\left[2\sigma_{i}^{-(l)}\rho\sigma_{i}^{+(l)}- \sigma_{i}^{+(l)}\sigma_{i}^{-(l)}\rho-\rho\sigma_{i}^{+(l)}\sigma_{i}^{-(l)}\right], \\ 
\mathcal{L}_{loss,\, cav}&=&\frac{\Gamma_{loss}}{2}\sum_{i=1}^{k}\left[2a_{i}\rho a_{i}^{\dagger}-a_{i}^{\dagger}a_{i}\rho-\rho a_{i}^{\dagger}a_{i}\right]
\end{eqnarray}
describing the pumping of the atoms, the spontaneous decay of the atoms, and the photon losses, respectively. The $\sigma_{i}^{\pm(l)}$ operators are the usual raising and lowering operators for the $l$-th atom in the $i$-th cavity. We introduce the detuning $\delta=\omega_{cav}-\omega_{at}$ of the bare cavity frequency with respect to the atomic frequency. In the following, we shall concentrate on a regime in which pumping of the atoms is much faster than their spontaneous decay, $\Gamma_{pump}\gg \gamma$, so the $\mathcal{L}_{loss,\, at}$ Lindblad term can be safely neglected. 

For simplicity, we will also restrict our attention to the $\Gamma_{pump}\gg\sqrt{N_{at}}\Omega_{R}$ regime, where the atoms are immediately repumped to their excited state after emitting a photon into the cavity: under such an assumption, an atom having decayed to the ground state does not have the time to reabsorb any photon before being repumped to its excited state. In this regime, complex cavity-QED effects such as Rabi oscillations do not take place and the photon emission takes place in an effectively irreversible way~\cite{Breuer,QuantumNoise}: as a result, we are allowed to eliminate the atomic dynamics from the problem and write a much simpler photonic master equation involving only the {cavity degrees of freedom}. 
%Nevertheless, a trace of the internal structure of the {atoms} will remain in the energy selectivity of the emission {processes}: photons will in fact be easier emitted into the cavity if the energy difference of the many photon states after and before emission is close to $\omega_{at}$. This feature is the key element of our proposal to selectively generate many-body states in the cavity lattice.

\subsection{Closed master equation for the {photonic} density matrix}\label{sec:Closed-master-equation}

Under the considered $\Gamma_{pump}\gg\Omega_R$ approximation, the atomic population is concentrated in the excited state and it is possible to use projective methods to write a closed master equation for the photonic density matrix where the atomic degrees of freedom $\mathcal{B}$ have been traced out, $\rho_{ph}=Tr_{\mathcal{B}}\rho$. All details of the (quite cumbersome) calculations can be found in \ref{app:projective}. The resulting photonic master equation reads 
\begin{equation}
\partial_{t}\rho_{ph}  =  -i\left[H_{ph},\rho_{ph}(t)\right]+\mathcal{L}_{loss} + \mathcal{L}_{em},
\label{eq:photon_only}
\end{equation}
with 
\begin{eqnarray}
\mathcal{L}_{loss} & = & \frac{\Gamma_{loss}}{2}\sum_{i=1}^{k}\left[2a_{i}\rho a_{i}^{\dagger}-a_{i}^{\dagger}a_{i}\rho-\rho a_{i}^{\dagger}a_{i}\right],\label{eq:loss}\\
\mathcal{L}_{em} & = & \frac{\Gamma_{em}}{2}\sum_{i=1}^{k}\left[\tilde{a}_{i}^{\dagger}\rho a_{i}+a_{i}^{\dagger}\rho\tilde{a}_{i}-a_{i}\tilde{a}_{i}^{\dagger}\rho-\rho\tilde{a}_{i}a_{i}^{\dagger}\right]. \label{eq:gainmarkov}
\end{eqnarray}
describing photonic losses and emission processes, respectively.
While the loss term has a standard Lindblad form at rate $\Gamma_{loss}$, the emission term keeps some memory of the atomic dynamics as it involves modified lowering and raising operators 
\begin{eqnarray}
\tilde{a}_{i}&=&\frac{\Gamma_{pump}}{2}\int_{0}^{\infty}d\tau\, e^{(-i\omega_{at}-\Gamma_{pump}/2)\tau}a_{i}(-\tau) ,
\label{eq:atilde} \\
\tilde{a}_{i}^{\dagger}&=&\left[\tilde{a}_{i}\right]^{\dagger} \label{eq:adagtilde}
\end{eqnarray}
which contain the photonic (hamiltonian and dissipative) dynamics during pumping. In the limit we are considering in which photonic losses are slow with respect to atomic pump, these operators are the interaction picture ones with respect to the photonic hamiltonian in the cavity array  and have a simpler expression~:
\begin{equation}
a_{i}(\tau)=e^{iH_{ph}\tau}\,a_{i}\,e^{-iH_{ph}\tau}.
\end{equation}
%Indeed, under the assumption $\Gamma_{pump}\gg\Gamma_{loss}$, one can say that in the integration kernel,ie over the time of atomic repumping, the photonic dynamic is purely hamiltonian
The Fourier-like integral in Eqs.\ref{eq:atilde} and \ref{eq:adagtilde} is responsible for the frequency selectivity of the emission, as the integral is maximum when the free evolution of $a_i$ occurs at a frequency close to the atomic one $\omega_{at}$. 

A deeper physical insight on the operators (\ref{eq:atilde}) and (\ref{eq:adagtilde}) can be obtained by looking at their matrix elements in the basis of eigenstates of the photonic hamiltonian. We consider two eigenstates $\ket f$ (resp. $\ket{f'}$) with $N$ (resp. $N+1)$ photons and energy $\omega_{f}$ (resp. $\omega_{f'}$). After elementary manipulation, we see that the emission amplitude follows a Lorentzian law as a function of the detuning between the frequency difference of the two photonic states $\omega_{f'f}=\omega_{f'}-\omega_{f}$ and the atomic transition frequency $\omega_{at}$,
\begin{equation}
%\bra f\tilde{a_{i}}\ket{f'} & = & \frac{\Gamma_{pump}/2}{i(\omega_{at}-\omega_{f'f})+\Gamma_{pump}/2}\bra fa_{i}\ket{f',}\label{eq:anni}\\
\bra{f'}\tilde{a}_{i}^{\dagger}\ket f = \frac{\Gamma_{pump}/2}{-i(\omega_{at}-\omega_{f'f})+\Gamma_{pump}/2}\bra{f'}a_{i}^{\dagger}\ket f~. \label{eq:crea}
\end{equation}
Upon insertion of Eq.\ref{eq:crea} into the master equation Eq.\ref{eq:photon_only}, one can associate the real part of the Lorentzian factor to an effective emission rate
\begin{equation}
\Gamma_{em}(\omega_{f'f})=\Gamma^0_{em} \frac{\Gamma^2_{pump}/4}{(\omega_{at}-\omega_{f'f})^2+\Gamma_{pump}^2/4},
\end{equation}
while the imaginary part can be related to a frequency shift of the photonic states under the effect of the population-inverted atoms. In the next section, this point will be made more precise under a secular approximation.

The width of the Lorentzian is set by the pumping rate $\Gamma_{pump}$, that is by the autocorrelation time $\tau_{pump}=1/\Gamma_{pump}$ of the atom seen as a frequency-dependent emission bath. The peak emission rate exactly on resonance is equal to
\begin{equation}
\Gamma^0_{em}=\frac{4N_{at}\Omega_{R}^{2}}{\Gamma_{pump}}\,.
\end{equation}
While the $\Gamma_{pump}\gg\sqrt{N_{at}}\Omega_R$ assumption automatically implies that the emission is much slower than the atomic repumping rate, $\Gamma_{em}\ll\Gamma_{pump}$, no constraint need being imposed on the parameters $J$, $U$ and $\delta=\omega_{cav}-\omega_{at}$ of the photonic Hamiltonian, which can be arbitrarily large. Whereas an extension of our study to the $\Gamma_{loss}\gtrsim \Gamma_{pump}$ regime would only introduce technical complications, entering the $\Gamma_{em}\gtrsim \Gamma_{pump}$ regime is expected to dramatically modify the physics, as a single atom could exchange photons with the cavity at such a fast rate that it has not time to be repumped to the excited state in between two emission events. As a result, reabsorption processes and Rabi oscillations are possible, which considerably complicate the theoretical description. {These issues will be the subject of future investigations.}

% We are interested in the stationary state 
% \begin{eqnarray*}
% 0 & = & -i\left[H_{ph},\rho\right]+\mathcal{L}_{loss,\, cav}\left(\rho\right)+\mathcal{L}_{em,\, cav}\left(\rho\right)
% \end{eqnarray*}
\subsection{Reformulation in Lindblad form in the secular approximation\label{sec:photonic-lindblad-form}}
In the case the system has a discrete spectrum, it is possible in the so-called secular approximation to write another photonic master implementing non-markovian effects with a more standard Lindblad form, compatible with Monte Carlo wave-function simulations \cite{Castin} and giving equivalent driven-dissipative dynamics. This can be explained by the following argument: in a weak dissipation limit ($\Gamma_{em},\,\Gamma_{loss}$ very small with respect to the gaps in the spectrum) terms of the density matrix $\rho_{f,\tilde{f}},\, \rho_{f',\tilde{f}'}$ which would be rotating at different frequencies $\omega_{f,\tilde{f}},\, \omega_{f',\tilde{f}'}$ if the system were isolated, are not coupled to each other by dissipation since the coupling $\Gamma^{0}_{em},\, \Gamma_{loss}$ is negligible with respect to their frequency difference $\Delta\omega=\omega_{f',\tilde{f}'}-\omega_{f,\tilde{f}}=\omega_{f',f}-\omega_{\tilde{f}',\tilde{f}}$. Considering this, all relevant dissipative transitions verify then $\Delta\omega\simeq 0$. Restricting the previous master equation given by Eqs.~\ref{eq:photon_only},\ref{eq:gainmarkov} and \ref{eq:crea} to these transitions, it is possible to rewrite the dynamics in the following way (details of the derivation are given in \ref{app:Lindblad-form}): 
\begin{equation}
\partial_{t}\rho_{ph}  =  -i\left[H_{ph}+\left(\sum_{i} H_{lamb,i}\right),\rho_{ph}(t)\right]+\mathcal{L}_{loss} + \bar{\mathcal{L}}_{em},
\label{eq:photon_only_MCWF}
\end{equation}
with 
\begin{equation}
\label{eq:gainmarkov_MCWF}
\bar{\mathcal{L}}_{em}(\rho_{ph})= \frac{\Gamma_{em}}{2}\sum_{i=1}^{k}\left[2\bar{a}_{i}^{\dagger}\rho_{ph} \bar{a}_{i}-\bar{a}_{i}\bar{a}_{i}^{\dagger}\rho_{ph}-\rho_{ph}\bar{a}_{i}\bar{a}_{i}^{\dagger}\right],
\end{equation}
\begin{equation}
\label{eq:crea_MCWF}
 \bra{f'}\bar{a}_{i}^{\dagger}\ket{f}=\frac{\Gamma_{pump}/2}{\sqrt{(\omega_{at}-\omega_{f',f})^2+\left(\Gamma_{pump}/2\right)^2}}\bra{f'}a_{i}^\dagger\ket{f},
\end{equation}
\begin{equation}
\label{eq:lamb_MCWF}
 \bra{f'}H_{lamb,i}\ket{f}=\sum_{f''} \bra{f'}a_{i}\ket{f''}\left(\frac{(\omega_{f'',f}-\omega_{at})\Gamma_{pump}/2}{(\omega_{at}-\omega_{f'',f})^2+(\Gamma_{pump}/2)^2}\right)\bra{f''}a_{i}^\dagger\ket{f}.
\end{equation}
Note that the jump operators $\bar{a}_i^\dagger$ have the same form as the ones considered in~\cite{Kapit} and have for effect to modify the the transition rate, while the "imaginary part" of Eq.~\ref{eq:crea} induces an additional Hamiltonian contribution in the form of a Lamb shift. Notice that the two master equations Eqs.~\ref{eq:photon_only},~\ref{eq:photon_only_MCWF} are slightly different. However, under the considered approximation they are expected to provide equivalent dynamics. The latter form has the advantage of being of Lindblad form, and thus is directly compatible with MCWF simulations \cite{Castin} and can be useful from a numerical point of view.

The secular approximation can be very restrictive (particularly in the thermodynamic limit where the spectrum is continuous). However, our feeling is that the reformulation of Eq.\ref{eq:photon_only_MCWF} should be accurate in a wider range of parameters. Quantitatively, we anticipate the condition $\Gamma_{em},\, \Gamma_{loss}\ll\Gamma_{pump}$ to be sufficient. More investigations in this direction are under way.
\section{One cavity}
\label{sec:onecavity}

As a first example of application, we consider the simplest case of a single nonlinear cavity.  A special attention will be paid to the stationary state $\rho_{ss}$ of the system for which Eq.\ref{eq:photon_only} imposes 
\begin{equation}
0=-i\left[H_{ph},\rho_{ss}\right]+\mathcal{L}_{loss}\left(\rho_{ss}\right)+\mathcal{L}_{em}\left(\rho_{ss}\right).
\end{equation}
In our specific case of a single cavity, the photonic states are labelled by the photon number $N$ and have an energy 
\begin{equation}
\omega_{N}=N\omega_{cav}+\frac{1}{2}N(N-1)U.
 \end{equation}
Correspondingly, the $N\rightarrow N+1$ transition has a frequency
\begin{equation}
\omega_{N+1,N}=\omega_{cav}+NU,
\end{equation}
and the corresponding photon emission rate is
\begin{equation}
\Gamma_{em}(\omega_{N+1,N})=\Gamma_{em}^0\frac{(\Gamma_{pump}/2)^{2}}{(\omega_{N+1,N}-\omega_{at})^{2}+(\Gamma_{pump}/2)^{2}}.
\end{equation}
As no coherence can exist between states with different photon number $N$, the stationary density matrix is diagonal in the Fock basis, $\rho_{ss}=\delta_{N,N'}\pi_N$ with the populations $\pi_N$ satisfying 
\begin{equation}
(N+1)\Gamma_{loss}\pi_{N+1}-(N+1)\Gamma_{em}(\omega_{N+1,N})\pi_{N}+N\Gamma_{em}(\omega_{N,N-1})\pi_{N-1}-N\Gamma_{loss}\pi_{N}=0,
\end{equation}
where the two last terms of course vanish for $N=0$. 
As only states with neighboring $N$ are connected by the emission/loss processes, detailed balance is automatically enforced in the stationary state, which imposes the simple condition on the populations,
\begin{equation}
(N+1)\Gamma_{loss}\pi_{N+1}-(N+1)\Gamma_{em}(\omega_{N+1,N})\pi_{N}=0
\end{equation}
%
%\begin{equation}
%(N+1)\tilde{\gamma}_{loss}\pi_{N+1}-(N+1)\tilde{\gamma}_{em}(\tilde{\omega}_{N+1,N})\pi_{N}=0.
%\end{equation}
which is straightforwardly solved in terms of a product,
\begin{equation}
\pi_{N} = {\pi_{0}}\,\prod_{M=0}^{N-1}\frac{\Gamma_{em}(\omega_{M+1,M})}{\Gamma_{loss}}= \left(\frac{\Gamma_{em}^0}{\Gamma_{loss}}\right)^{N}\prod_{M=0}^{N-1}\frac{(\Gamma_{pump}/2)^{2}}{(\omega_{M+1,M}-\omega_{at})^{2}+(\Gamma_{pump}/2)^{2}}\pi_{0}.
\end{equation}
The excellent agreement of this result with a numerical solution of the full atom-cavity system is illustrated in \ref{app:valid}.

\subsection{Linear regime}

\begin{figure*}[htbp]
\begin{center}
 \includegraphics[height=0.40\textwidth,clip]{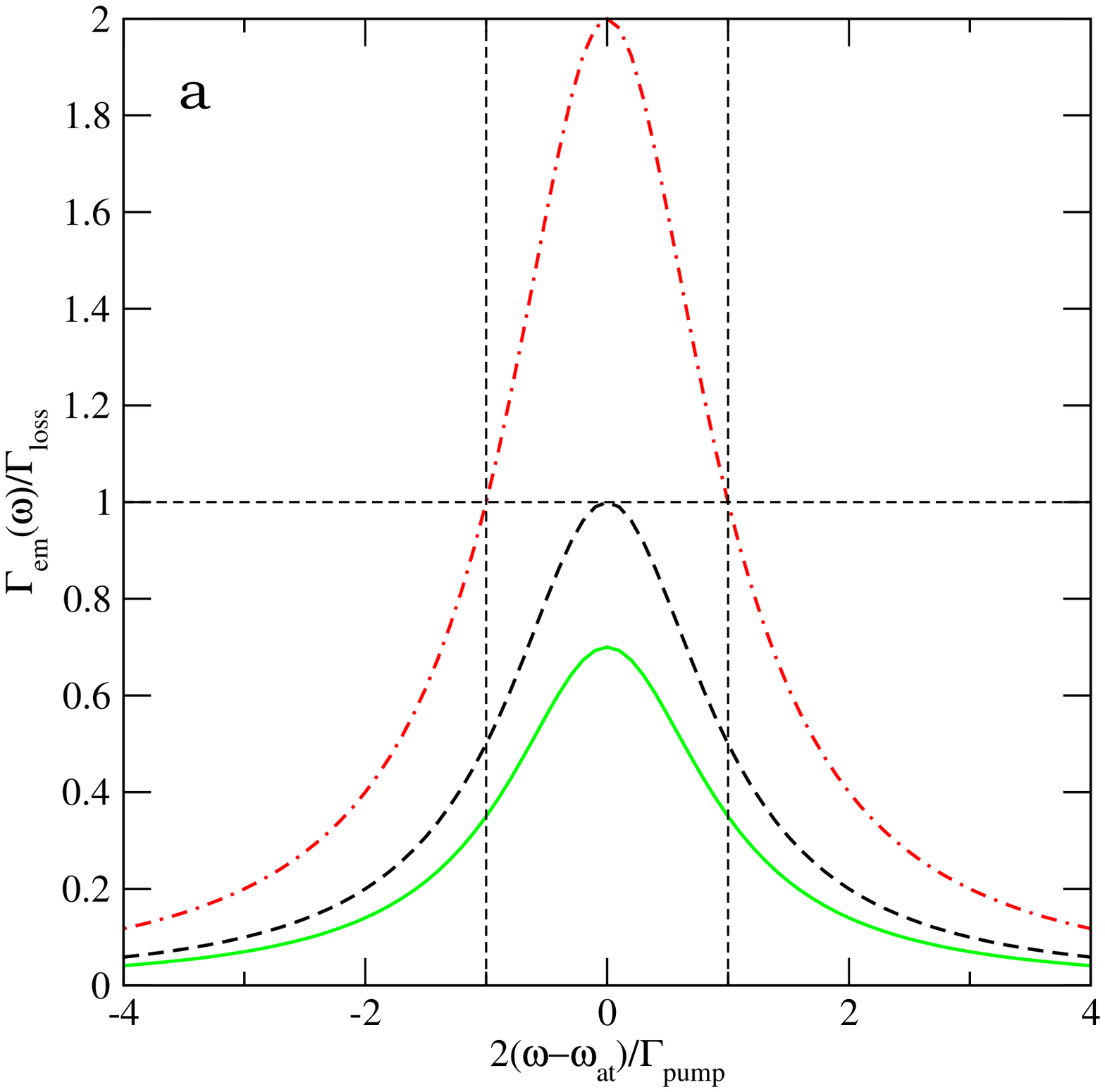}\hspace{1cm}
\includegraphics[height=0.40\textwidth,clip]{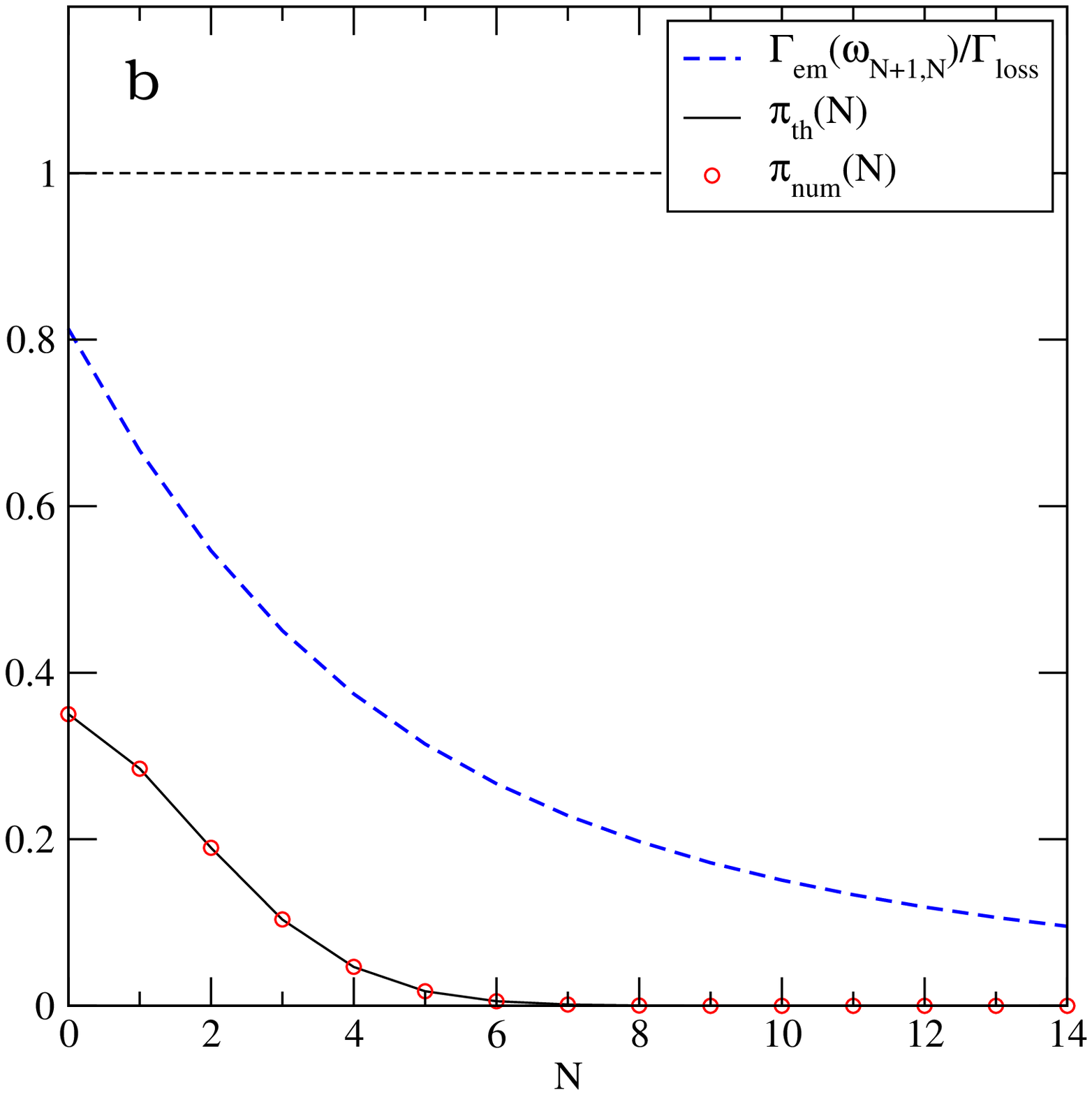}\\ \vspace*{1.5cm}
\includegraphics[height=0.40\textwidth,clip]{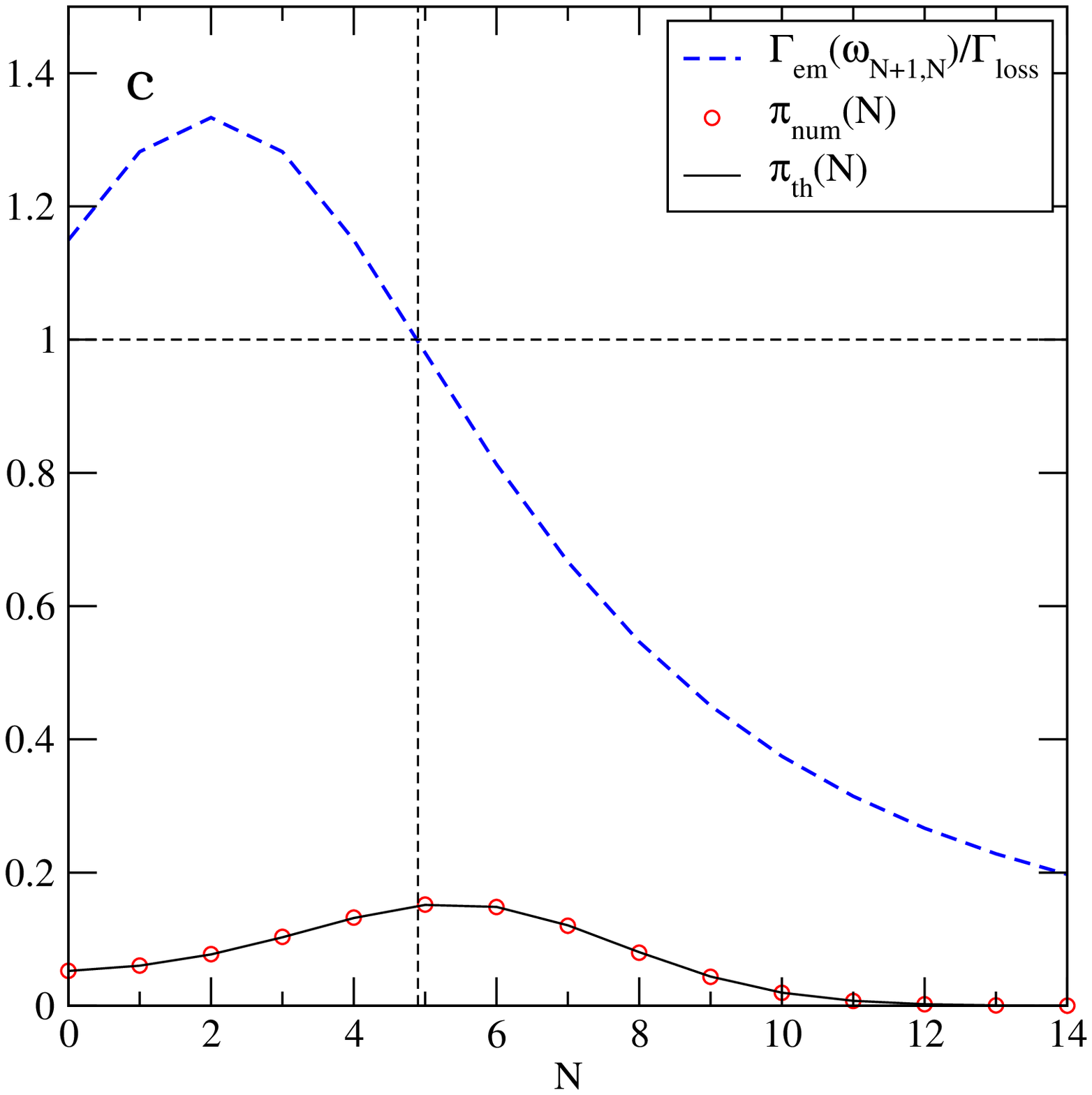} \hspace{1cm}
\includegraphics[height=0.40\textwidth,clip]{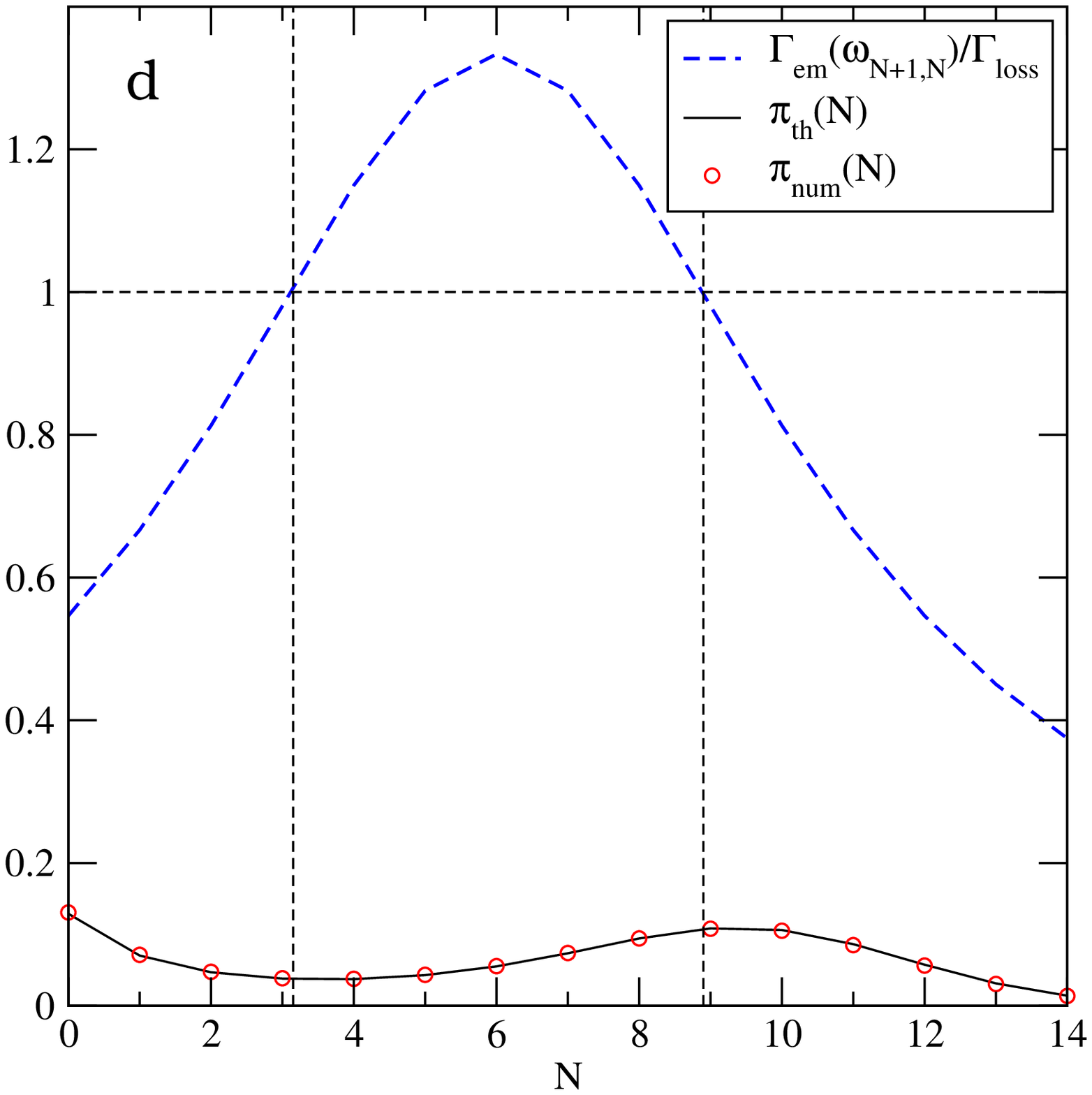}\\ \vspace*{1.cm}
\end{center}
\caption{(a) Emission vs. loss rate as a function of the detuning from the atomic frequency $\omega_{at}$: the three curves are for peak emission $\Gamma_{em}^0$ larger ({red dash-}dotted), equal ({black} dashed), smaller ({green} solid) than the loss rate $\Gamma_{loss}$.
(b-d) Populations $\pi_N$ of the $N$-photon state as a function of $N$ in the three cases $\omega_{2}\leq\omega_{cav}$ (b), $\omega_{1}\leq\omega_{cav}\leq\omega_{2}$ (c), $\omega_{cav}\leq\omega_{1}$ (d). In the three panels, the open dots are the numerical results of the atom-cavity theory, while the {solid} line is the prediction of the analytical purely photonic theory; the {dashed} curves show the ratio $\Gamma_{em}(\omega_{N+1,N})/\Gamma_{loss}$ as a function of $N$.
Parameters: $\delta/U=4$ (b), $-2$ (c), $-6$ (d). In all panels,  $2U/\Gamma_{pump}=0.2$, 
 $2\Gamma_{loss}/\Gamma_{pump}=0.0006$, 
 $2\Omega_{R}/\Gamma_{pump}=0.02$. 
 \label{fig:Gain-and-populations}}
\end{figure*}

For a vanishing nonlinearity $U=0$, all transition frequencies $\omega_{N+1,N}$ are equal to the bare cavity frequency $\omega_0$ and the {populations} of the different $N$ states have a constant ratio
\begin{equation}
\frac{\pi_{N+1}}{\pi_{N}}=\frac{\Gamma_{em}^0}{\Gamma_{loss}}\frac{(\Gamma_{pump}/2)^{2}}{\delta^{2}+(\Gamma_{pump}/2)^{2}},
\end{equation}
where we remind that $\delta=\omega_{cav}-\omega_{at}$.
%\begin{equation}
%\frac{\pi_{N+1}}{\pi_{N}}=\frac{\tilde{\gamma}_{em,0}}{\tilde{\gamma}_{loss}}\frac{1}{\tilde{\omega_0}^{2}+1}.
%\end{equation}
For weak pumping %emission ?
and/or large detuning, {one has}
\begin{equation}
\Gamma_{em}^0\frac{(\Gamma_{pump}/2)^{2}}{\delta^{2}+(\Gamma_{pump}/2)^{2}}<\Gamma_{loss} ,
\end{equation}
{so} the density matrix for the cavity shows a monotonically decreasing thermal occupation law. 
For strong pumping %emission ?
and close to resonance{, one can achieve the regime where the emission overcompensates} losses and the cavity mode starts being strongly populated: 
 \begin{equation}
\Gamma_{em}^0\frac{(\Gamma_{pump}/2)^{2}}{\delta^{2}+(\Gamma_{pump}/2)^{2}}>\Gamma_{loss}  .
 \end{equation}
The transition between the two regimes is the usual laser threshold, but our purely photonic theory is not able to include the gain saturation mechanism that serves to stabilize laser oscillation above threhsold~\cite{laser,QuantumNoise}: within our purely photonic theory, the population {would in fact show} a clearly unphysical monotonic growth for increasing $N$. A complete description in terms of the full atom-cavity master equation would of course solve this pathology including a gain saturation mechanism according to usual laser theory{, but this goes beyond the scope of the present work.}

\subsection{Optical bistability phenomena in weak nonlinear cavities}
For $U>0$, the situation is much more interesting as the effective transition frequency depends on the number of photons, 
\begin{equation}
\omega_{N+1,N}=\omega_{cav}+NU\geq\omega_{cav}, \hspace{1cm},
\end{equation}
so the gain condition
\begin{equation}
\frac{\Gamma_{em}^0}{\Gamma_{loss}}\frac{(\Gamma_{pump}/2)^{2}}{(\omega_{N+1,N}-\omega_{at})^{2}+(\Gamma_{pump}/2)^{2}}\geq1\label{eq:lasing}
\end{equation}
can be satisfied in a finite range of photon numbers only, as it is illustrated in Fig.\ref{fig:Gain-and-populations}(a). As a consequence, even a weak nonlinearity $U$ is able to stabilize the system for any value of $\Gamma_{em}^0$ even in the absence of any gain saturation mechanism.

For $\Gamma_{em}^0<\Gamma_{loss}$, losses always dominate. For $\Gamma_{em}^0>\Gamma_{loss}$, the gain condition is instead satisfied in a range of frequencies $[\omega_1,\omega_2]$ around $\omega_{at}$. Under the weak nonlinearity condition $U\ll \Gamma_{pump}$, the $[\omega_1,\omega_2]$ range typically contains a large number of transition frequencies $\omega_{N+1,N}$ at different $N$. Three different regimes can then be identified depending on the position of the cavity frequency $\omega_{cav}$ with respect to the $[\omega_1,\omega_2]$ range.
 
(i) If $\omega_{2}\leq\omega_{cav}$, then the gain condition is never verified, and the population $\pi_{N}$ shown in Fig.\ref{fig:Gain-and-populations}(b) is a monotonically decreasing function of $N$. In this regime, the state of the cavity field is very similar to a thermal state, as it usually happens in a laser below threshold. (ii) If $\omega_{1}\leq\omega_{cav}\leq\omega_{2}$, the population $\pi_N$ shown in Fig.\ref{fig:Gain-and-populations}(c) is an increasing function for small $N$, shows a single maximum for $N\simeq \bar{N}=(\omega_2-\omega_{cav})/U$, and finally monotonically decreases for $N>\bar{N}$.

\begin{figure}[htbp]
\begin{center}
\includegraphics[width=0.4\textwidth,clip]{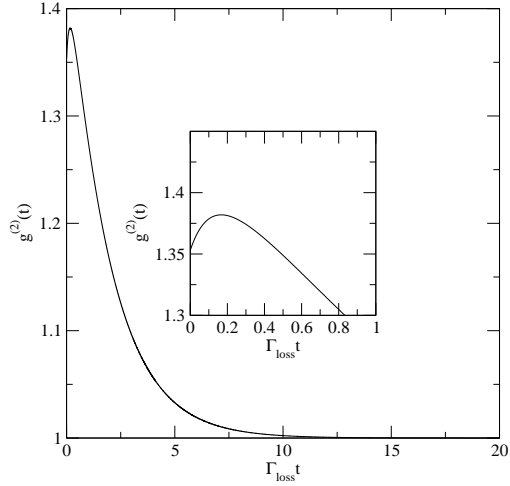}
\end{center}
\caption{
\label{fig:g2_one_weak}
Purely photonic simulation of the two-time coherence function $g^{(2)}(\tau)$ in the weakly nonlinear regime.
Parameters $U/\Gamma_{pump}=0.1$, $\Gamma_{loss}/\Gamma_{pump}=0.03$, $\Gamma^0_{em}/\Gamma_{pump}=0.04$, $\delta=-6U$ as in Fig.\ref{fig:Gain-and-populations}(d).
}
\end{figure}

The phenomenology is the richest in the regime (iii) where $\omega_{cav}\leq\omega_{1}$. In this case, for small $N$ the population $\pi_N$ decreases from its initial value $\pi_0$ until the nonlinearly shifted frequency enters in the gain interval for $N\simeq \bar{N}'=(\omega_1-\omega_{cav})/U$. After this point $\pi_N$ starts increasing again until it reaches a local maximum at $N\simeq \bar{N}=(\omega_2-\omega_{cav})/U$. Finally, for even larger $N$ it begins to monotonically decrease. An example of this complicate behaviour is shown in Fig.\ref{fig:Gain-and-populations}(d).

The existence of two well separate local maxima at $N=0$ and $N\simeq \bar{N}$ in the photon number distribution $\pi_N$ suggests that the incoherently driven nonlinear cavity exhibits a sort of bistable behaviour: when it is prepared at one maximum of the photon number distribution $\pi_N$, the system is trapped in a metastable state localized in a neighborhood of this maximum for a macroscopically long time. Switching from one metastable state to the other results is only possible as a result of a large fluctuation, so it has a very low probability, typically exponentially small in the photon number difference between the two metastable states.

This bistable behavior is clearly visible in the temporal dependence of the delayed two-photon correlation function
\begin{equation}
g^{(2)}(\tau)=\frac{\langle a^{\dagger}(t)\,a^{\dagger}(t+\tau)\,a(t+\tau)\,a(t) \rangle_{ss}}{\langle a^{\dagger}(t)\,a (t) \rangle_{ss}\langle a^{\dagger}(t+\tau)\,a(t+\tau) \rangle_{ss}}:
\end{equation} 
that is plotted in Fig.\ref{fig:g2_one_weak}. At short times, the value of $g^{(2)}$ is determined by a weighted average of the contribution of the two maxima according to the stationary $\pi_N$.  After a quick transient of order $1/\Gamma_{em,loss}$, which corresponds to a fast local equilibration of the probability distribution around each of its maxima, the $g^{(2)}$ correlation function slowly decays to its asymptotic value $1$ on a much longer time-scale mainly set by the exponentially long switching time from one maximum to the other

Before proceeding, it is worth emphasizing that the present mechanism for optical bistability bears important differences from the dispersive or absorptive optical bistability phenomena discussed in textbooks~\cite{Boyd,B-C}. On one hand there is some analogy to dispersive optical bistability in that the intensity-dependence of the refractive index is responsible for a frequency shift of the cavity resonance; on the other hand the frequency-selection is not provided by the resonance condition with a monochromatic coherent incident field rather by the frequency dependence of the gain due to the incoherent pump.

\subsection{Photon number selection in strongly nonlinear cavities \label{sub:Strongly-nonlinear-cavity,}}

In the opposite limit $U\gg\Gamma_{pump}$, the nonlinearity is so large that a change of photon number by a single unity has a sizable effect on the emission rate $\Gamma_{em}(\omega_{N+1,N})$. As discussed in \ref{app:projective}, the derivation of the photonic master equation remains fully valid in this regime provided $\Gamma_{pump}\gg\Gamma_{em}^0,\Gamma_{loss}$.

The ensuing physics is most clear in the regime when the maximum emission rate is large but only a single transition fits within the emission lineshape: these assumptions are equivalent to imposing that 
\begin{equation}
\frac{\Gamma_{em}^0}{\Gamma_{loss}}\gg 1\hspace{1cm}\textrm{and}\hspace{1cm} \frac{\Gamma_{em}^0}{\Gamma_{loss}}\frac{\Gamma_{pump}^2}{U^{2}}\ll1
\end{equation}
with the further condition that the emission is resonant with the $N_0\rightarrow N_0+1$ transition, 
\begin{equation}
\omega_{at}=\omega_{cav}+N_{0}U. 
\end{equation}
As a result, only this last transition is dominated by emission, while all others are dominated by losses.

\begin{figure}
\begin{center}
\includegraphics[scale=0.3,clip]{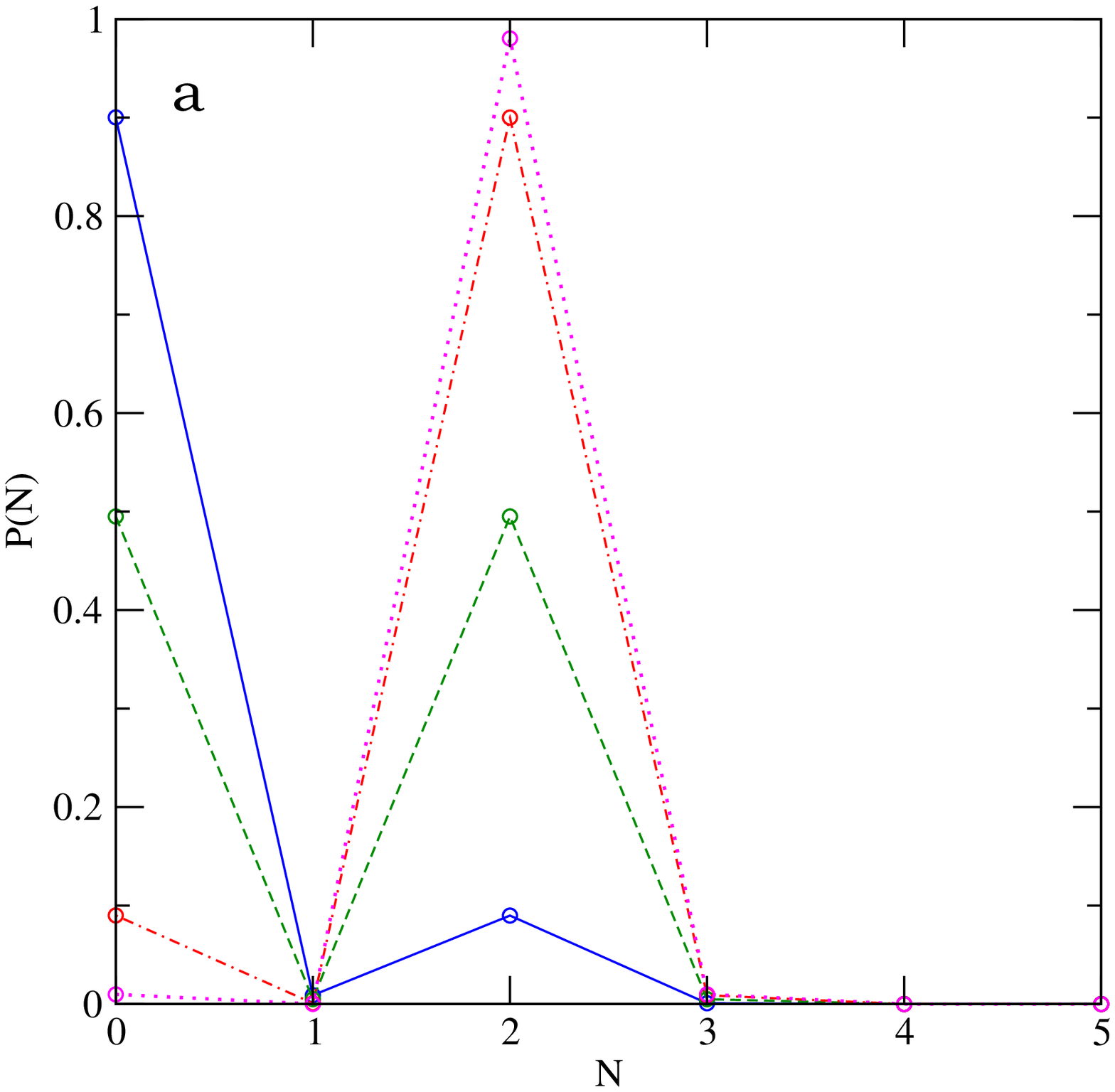} 
\hspace{1cm}\includegraphics[scale=0.3,clip]{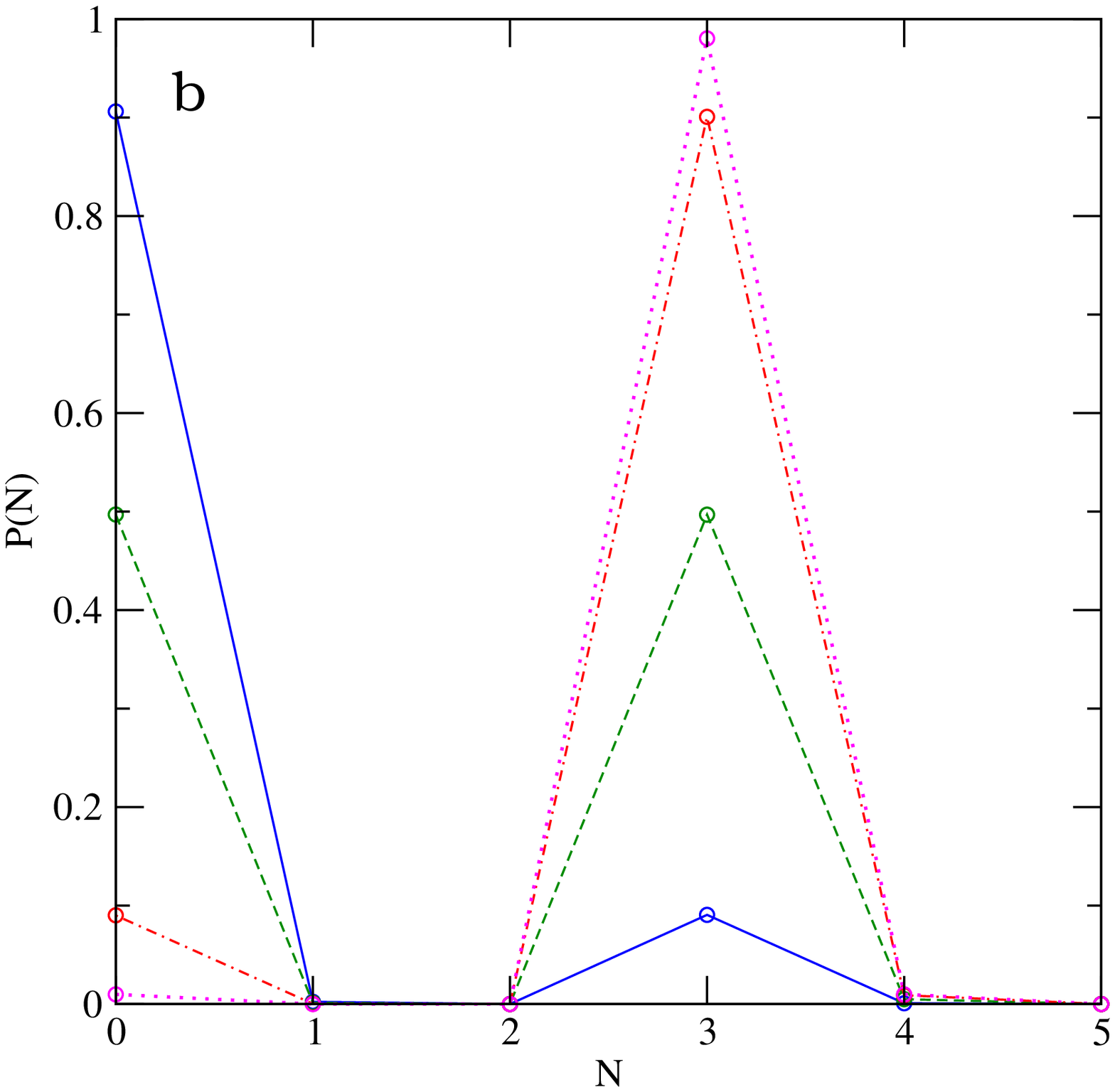}
\end{center}
\caption{
{Selective generation of a $N_0=2$ photon (upper panel) and $N_0=3$ photon (lower panel) Fock state: Population $\pi_N$ as a function of $N$ for different pumping parameters. The points are the result of a purely photonic simulation, the lines are a guide to the eye.
Left panel parameters:} for all curves $\delta=-U$, $2\Omega_R/\Gamma_{pump}=0.01$, and then for each particular curve $2\Gamma_{loss}/\Gamma_{pump}=2\,10^{-5}$ (blue solid line), $2\,10^{-6}$ (green, dashed line), $2\,10^{-7}$ (red, dash-dotted line), $2\,10^{-8}$ (magenta, dotted line). $2U/\Gamma_{pump}=10^{3/2}$ (blue solid line), $10^{2}$ (green, dashed line), $10^{5/2}$ (red, dash-dotted line), $10^{3}$ (magenta, dotted line).
{Right panel parameters:} fora ll curves $\delta=-U$, $2\Omega_R/\Gamma_{pump}=0.01$, and then $2\Gamma_{loss}/\Gamma_{pump}=5\,10^{-8}$ (blue solid line), $5\,10^{-9}$ (green, dashed line), $5\,10^{-10}$ (red, dash-dotted line), $5\,10^{-11}$ (magenta, dotted line). $2U/\Gamma_{pump}=2\,10^{5/2}$ (blue solid line), $2\,10^{3}$ (green, dashed line), $2\,10^{7/2}$ (red, dash-dotted line), $2\,10^{4}$ (magenta, dotted line).
The goal of {these} {choices} of parameters was to control the steady-state ratios $P(N+1)/P(N)=10^{-2}$ and $P(N)/P(0)=0.1, 1, 10, 100$ (blue, green, red, magenta).
\label{fig:one_strongly}}
\end{figure}

\begin{figure}[htbp]
\begin{center}
\includegraphics[width=0.35\textwidth,clip]{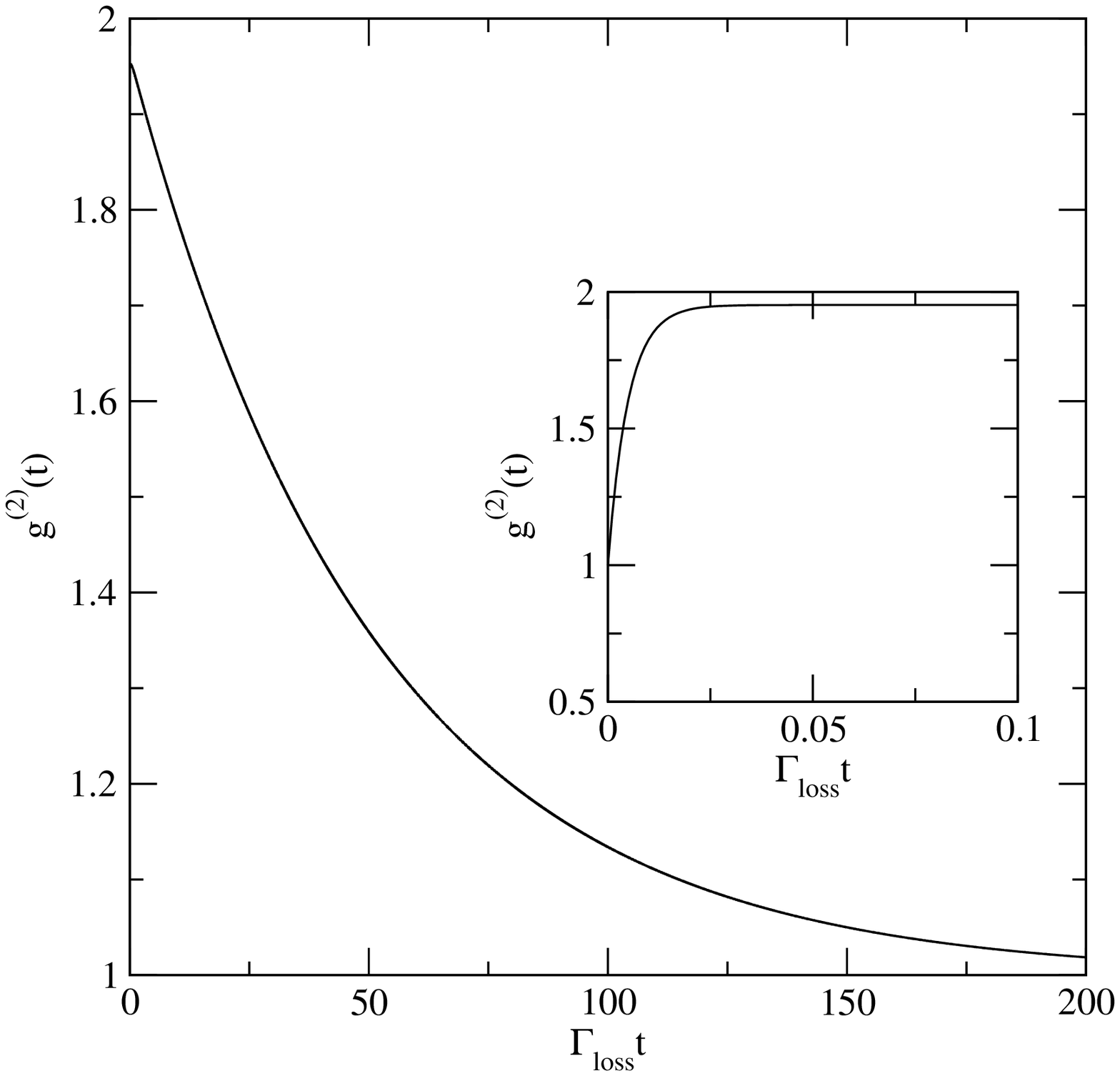}\hspace{1cm}
\includegraphics[width=0.35\textwidth,clip]{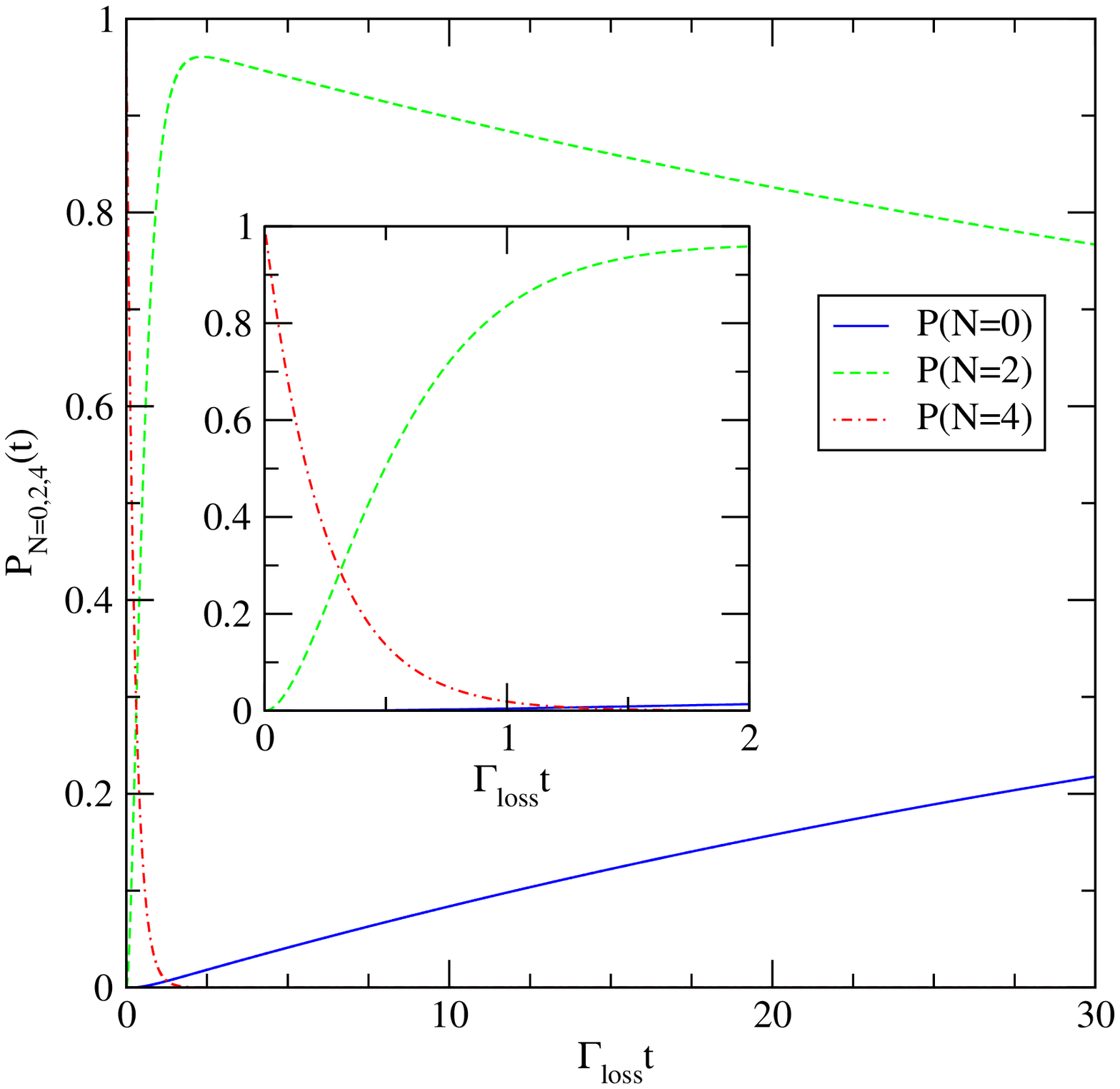}
\end{center}
\caption{
\label{fig:g2_one_strongly_generation}
Left panel: Purely photonic simulation of the two-time coherence function $g^{(2)}(\tau)$ for a strongly nonlinear regime in a (metastable) $N_0=2$ photon selection regime. The inset shows a magnified view of the short time region.
Parameters: $2U/\Gamma_{pump}=100$, $2\Gamma_{loss}/\Gamma_{pump}=2\, 10^{-3}$, $2\Gamma_{em}/\Gamma_{pump}=0.2$, $\delta=-U$; in the language of Fig.\ref{fig:one_strongly}, the present parameters would correspond to a regime where the $N=0,2$ states are almost equally occupied.
Right panel:  Preparation of the metastable state at $N_0=2$ starting from a $N=4$ %$N=5$ initial state. The different curves show the evolution in time of 
$\pi(4)$ (red {dot-dashed}) $\pi(2)$ (green {dashed}) $\pi(0)$ (blue {solid}). Same parameters as in Fig.\ref{fig:one_strongly}.
%$2U/\Gamma_{pump}=100$, $2\Gamma_{loss}/\Gamma_{pump}=2\, 10^{-3}$, $2\Gamma_{em}/\Gamma_{pump}{=}0.2$, $\delta=-U$.
} 
\end{figure}

% \begin{figure}[htbp]
% \begin{center}
% \includegraphics[width=0.3\textwidth,clip]{two_point_correlation_strongly_interacting}
% \end{center}
% \caption{
% \label{fig:g2_one_strongly}
% Two-time coherence function $g^{(2)}(\tau)$ for a strongly nonlinear regime in the $N_0=2$ photon selection regime (metastable). The inset shows a magnified view of the short time region.
% Parameters: $2U/\Gamma_{pump}=100$, $2\Gamma_{loss}/\Gamma_{pump}=2\, 10^{-3}$, $2\Gamma_{em}/\Gamma_{pump}=0.2$, $\delta=-U$.
% In the language of Fig.\ref{fig:one_strongly}, the present parameters would correspond to a regime where the $N=0,2$ states are almost equally occupied.} 
% \end{figure}
% 
% \begin{figure}[htbp]
% \includegraphics[width=0.48\textwidth,clip]{generation_pop.eps}
% \caption{
% \label{fig:generation_pop}
% Preparation of the metastable state at $N_0=2$ starting from a $N=4$ %$N=5$
%  initial state. The different curves show the evolution in time of 
% $\pi(4)$ (red {dot-dashed}) $\pi(2)$ (green {dashed}) $\pi(0)$ (blue {solid}). Same parameters as in Fig.\ref{fig:one_strongly}.
% %$2U/\Gamma_{pump}=100$, $2\Gamma_{loss}/\Gamma_{pump}=2\, 10^{-3}$, $2\Gamma_{em}/\Gamma_{pump}{=}0.2$, $\delta=-U$.
% }
% \end{figure}

In terms of the diagrams in Fig.\ref{fig:Gain-and-populations}, the stationary distribution $\pi_N$ is therefore sharply peaked at two specific values, $N=0$ and at $N=N_0$. Examples of this physics are illustrated in Fig.\ref{fig:one_strongly}: the two peaks are always clearly visible, but depending on the parameters their relative height can be tuned to different values almost at will. It is however important to note that having a sizable stationary population in the $N=N_0$ peak requires quite extreme values of the parameters as population would naturally tend to accumulate at $N=0$ and this difficulty turns out to be exponentially harder for larger $N_0$.

The physics underlying this behaviour can be easily explained in terms of the asymmetry in the switching mechanisms leading from $N=0$ to $N=N_0$ and viceversa. The former process requires in fact a sequence of several unlikely emission events from $N=0$ to $N=N_0-1$ as emission is favoured only in the last step. On the other hand, decay from $N=N_0$ occurs as a consequence a single unlikely loss event from $N=N_0-1$ to $N=N_0-2$: as soon as the system is at $N=N_0-2$, it will quickly decay to $N=0$. 

The rate $\Gamma_{acc}$ of such an accident can be estimated as follows: the probability that the system in $N=N_0-1$ decays to $N=N_0-2$ is a factor ${(N_{0}-1)\Gamma_{loss}}/({N_{0}\Gamma_{em}^0})$ smaller than the one of being repumped to $N=N_0$. As the rate at which the system decays from $N=N_0$ to $N_0-1$ is approximately equal to $N_{0}\Gamma_{loss}$, one finally obtains
\begin{equation}
\Gamma_{acc}=N_{0}\Gamma_{loss}\frac{(N_{0}-1)\Gamma_{loss}}{N_{0}\Gamma_{em}^0}\ll N_{0}\Gamma_{loss}.
\end{equation}
This longer time scale $\tau_{acc}=\Gamma_{acc}^{-1}$ is clearly visible in the long tail of the time-dependent $g^{(2)}(t)$ that is plotted in the left panel of Fig.\ref{fig:g2_one_strongly_generation}. The quick feature at very short times corresponds to the emission rate $\Gamma_{em}$.

If needed, the characteristic time scale $\tau_{acc}$ could be further enhanced by adding a second atomic species whose transition frequency is tuned to quickly and selectively emit photons on the $N-2\rightarrow N-1$ transition. In this way, the accident rate can be efficiently reduced to $\Gamma_{acc}^{(2)}\simeq\Gamma_{loss}\left(\Gamma_{loss}/\Gamma_{em}^0\right)^{2}\ll\Gamma_{acc}$. By repeating the mechanism on $k$ transitions, one can suppress the accident rate in a geometrical way to $\Gamma_{acc}^{(k)}\simeq\Gamma_{loss}\left(\Gamma_{loss}/\Gamma_{em}^0\right)^{k}\ll\Gamma_{acc}$. Finally, the Fock state with $N_0$ photons can be fully stabilized to an infinite lifetime and no problem of metastability if $N_0$ different atomic species are included so to cover all transitions from $N=0$ to $N=N_0$.

From a slightly different perspective, we can take advantage of {the slow rate of accidents $\Gamma_{acc}$} to selectively prepare a metastable state with $N=N_0$ photons even in parameter regimes where the $N=0$ state would be statistically favoured at steady-state. Though the state will eventually decay to $N=0$, the lifetime of the metastable $N=N_0$ state can be long enough to be useful for interesting experiments: The idea to prepare the state with $N_0$ photons is to inject a larger number $N>N_0$ of photons into the cavity: the system will quickly decay to the $N=N_0$ state where the system remains trapped with a lifetime $\Gamma_{acc}^{-1}$. 

The efficiency of this idea is illustrated in the right panel of Fig.\ref{fig:g2_one_strongly_generation} where we plot the time evolution of the most relevant populations $\pi_N$. The initially created state with $N=N_{in}$ photons quickly decays, so that population accumulates into $N=N_0$ on a time-scale of the order of $\Gamma_{loss}$; the eventual decay of the population towards $N=0$ will then occur on a much longer time set by $\Gamma_{acc}$. It is worth noting that this strategy does not require that the initial preparation be number-selective: it will work equally well if a wide distribution of $N_{in}$ are generated at the beginning, provided a sizable part of the distribution lies at $N>N_0$. Furthermore, this idea removes the need for extreme parameters such as the ones used in Fig.\ref{fig:one_strongly} to obtain a balance between $\pi(N)$ and $\pi(0)$: as a result, the difficulty of creating a (metastable) state of $N_0$ photons is roughly independent of $N_0$. 

{These} results show the potential of this novel photon number selection scheme to obtain light pulses with novel nonclassical properties: for instance, upon a sudden switch-off of the cavity mirrors, {one would obtain} a wavepacket containing an exact number of photons {sharing the same wavefunction}. With respect to the many other configurations discussed in the recent literature to produce $N$-photon Fock states and photon bundles~\cite{Majumdar,Rundquist,Munoz}, our proposal has the advantage of giving a deterministic preparation of a $N$-photon Fock state in the cavity, which can then be manipulated to extract light pulses with the desired quantum properties.

\section{Cavity arrays} \label{sec:Arrays-of-cavities}

After having unveiled a number of interesting features that occur in the simplest case of a single-cavity, we are now in a position to start attacking the far richer many-cavity case. From now on we consider that the isolated photonic Hamiltonian is the Bose-Hubbard one with tunelling $J$ and interaction constant $U$. Throughout this section, we shall make heavy use of the purely photonic description previously derived, which allows to consider bigger systems with a higher number of photons. A numerical validation of this approach against the solution of the full atom-cavity master equation is presented in \ref{app:valid}.

\subsection{Markovian regime} \label{sec:Markov}

We begin by considering the Markovian limit of the theory, which is recovered for $\Gamma_{pump}=\infty$, i.e. for a frequency-independent gain. In this case, the emission term of the master equation for photons Eq.\ref{eq:gainmarkov} reduces to the usual Lindblad form
\begin{equation}
\mathcal{L}_{em} =\frac{\Gamma_{em}^0}{2}\sum_{i=1}^{k}\left[2a_{i}^{\dagger}\rho a_{i}-a_{i}a_{i}^{\dagger}\rho-\rho a_{i}a_{i}^{\dagger}\right].
\end{equation}
For a single cavity, the stationary state is immediately obtained as 
\begin{equation}
\pi_{N}=\frac{1}{1-\frac{\Gamma_{em}^0}{\Gamma_{loss}}}\left(\frac{\Gamma_{em}^0}{\Gamma_{loss}}\right)^{N}:
\end{equation}
a necessary condition for stability for this system is of course that $\Gamma_{em}^0<\Gamma_{loss}$. For $\Gamma_{em}^0>\Gamma_{loss}$ amplification would in fact exceed losses and the system display a laser instability: while a correct description of gain saturation is beyond the purely photonic theory, the full atom-cavity theory would recover for this model the standard laser operation~\cite{QuantumNoise,Breuer,laser}.

For larger arrays of $k$ sites, a straightforward calculation shows that in the Markovian limit the stationary matrix keeps a structureless form, 
\begin{equation}
\rho_{\infty}=\sum_{N}\pi_{N}\mathcal{I}_{N} ,
\end{equation}
with 
\begin{equation}
\pi_{N}=\frac{1}{\sum_{M}D_{M}\left(\frac{\Gamma_{em}^0}{\Gamma_{loss}}\right)^{M}}\left(\frac{\Gamma_{em}^0}{\Gamma_{loss}}\right)^{N}.
\end{equation}
Here, $D_{N}=\frac{(N+k-1)!}{(k-1)!N!}$ is the dimension of the Hilbert subspace with a total number of photons equal to $N$ and $\mathcal{I}_{N}$ is the projector over this subspace. The interested reader can find the details of the derivation in \ref{app:Exact-stationary-solution}. 

This result shows that independently of the number of cavities and the details of the Hamiltonian, in the Markovian limit the density matrix in the stationary state corresponds to an effective Grand-Canonical ensemble at infinite temperature $\beta=0$ with a fugacity $z=e^{\beta\mu}={\Gamma_{em}^0}/{\Gamma_{loss}}$ determined by the pumping and loss conditions only: All states are equally populated and the system does not display much interesting physics. In particular, the steady state does not depend on the tunelling amplitude $J$ and on the photon-photon interaction constant $U$

%Maybe not so radical~: even for weakly non markovian we are limited to  U,J<<Gamma_pump=kT so the system is in a semiclassical limit~: the single particle do es not matter no condensation, no quantum effect%

\subsection{Effective Grand-Canonical distribution in a weakly non-Markovian and secular regime \label{sub:GC}}

The situation changes as soon as some non-Markovianity is included in the model. In this section we start from a weakly non-Markovian case where all relevant transitions adding one photon have a narrow distribution around the bare cavity frequency, $|\omega_{f'f}-\omega_{cav}|\ll\Gamma_{pump}$. We also assume a secular limit where $U,\, J\gg\Gamma^0_{em},\,\Gamma_{loss}$, so that the non-diagonal terms of the density matrix {in the photonic hamiltonian eigenbasis} oscillate at a fast rate and are thus effectively decoupled from the (slowly varying) populations. In this limit, we can safely assume that all coherences vanish and we can restrict our attention to the populations. This somehow critical approximation will be justified a posteriori in the next section, where we treat perturbatively the coupling of populations to coherences and show both analytically and numerically that in the weakly markovian regime, their contribution is of higher order in the 'non-markovianity' parameter $1/\Gamma_{pump}$ and therefore can be safely neglected.

Under these assumptions, the transfer rate on the $\ket{f'}\rightarrow \ket{f}$ transition where one photon is lost from $N+1$ to $N$ has a frequency-independent form
\begin{equation}
T_{f' \rightarrow f} = \Gamma_{loss}\left|\bra{f} a \ket{f'}\right|^{2},
\end{equation}
while the reverse emission process depends on the detunings $\Delta_{f'f}=\omega_{f'f}-\omega_{cav}$ and $\delta=\omega_{cav}-\omega_{at}$ as
\begin{equation}
 T_{f\rightarrow f'} = \Gamma_{em}^0\left|\bra{f'} a^\dagger \ket{f}\right|^2 \frac{\frac{\Gamma_{pump}^2}{4}}{(\Delta_{f'f}+\omega_{cav}-\omega_{at})^{2}+\frac{\Gamma_{pump}^2}{4}}
 \simeq  \tilde{\Gamma}_{em}^0\left|\bra fa^\dagger \ket{f'}\right|^2 \left[1-\beta\Delta_{f'f}+\mathcal{O}\left(\Delta_{f'f}\right)^{2}\right],
 \label{eq:emissonWMS}
\end{equation}
with 
\begin{eqnarray}
\tilde{\Gamma}_{em}^0 & = & \frac{\left(\Gamma_{pump}/2\right)^{2}}{(\omega_{cav}-\omega_{at})^{2}+\left(\Gamma_{pump}/2\right)^{2}}\Gamma_{em}^0 ,\\
\beta & = & \frac{2(\omega_{cav}-\omega_{at})}{(\omega_{cav}-\omega_{at})^{2}+\left(\Gamma_{pump}/2\right)^{2}}.
\label{eq:beta}
\end{eqnarray}
In this expression, the weakly non-Markovian regime is characterized by having $|\beta \Delta_{f'f} | \ll 1$: in this case, the square bracket in Eq.\ref{eq:emissonWMS} can be replaced with no loss of accuracy by an exponential
\begin{equation}
1-\beta\Delta_{f'f}\simeq e^{-\beta\Delta_{f'f}},
\end{equation}
which immediately leads to a Grand-Canonical form of the stationary density matrix 
\begin{equation}
\rho_{\infty}=\frac{1}{\Xi}e^{\beta N\mu}e^{-\beta H},
\end{equation}
with an effective chemical potential 
\begin{equation}
\mu=\frac{1}{\beta}\log\left(\frac{\tilde{\Gamma}_{em}^0}{\Gamma_{loss}}\right)+\omega_{cav} 
\end{equation}
and an effective temperature $k_B T=1/\beta$: most remarkably, even if each transition involves a small deviation from the bare cavity frequency $\omega_{cav}$, %in the thermodynamic limit
the cumulative effect of many such deviations can have important consequences for large photon numbers, so to make the stationary distribution strongly non-trivial. Remarkably, both positive and negative temperature configurations can be obtained from Eq.\ref{eq:beta} just by tuning the peak emission frequency $\omega_{at}$ either below or above the bare cavity frequency $\omega_{cav}$. 
As expected for a thermal-like distribution, detailed balance between eigenstates is satisfied
 \begin{multline}
T_{f'\rightarrow f}\pi_{f'}-T_{f\rightarrow f'}\pi_{f}=\left|\bra{f'} a^\dagger \ket{f}\right|^{2}  \,\left[\Gamma_{loss}\frac{1}{\Xi}\left(\frac{\tilde{\Gamma}_{em}^0}{\Gamma_{loss}}e^{\beta\omega_{cav}}\right)^{N+1}e^{-\beta \omega_{f'}} \right. +\\
-\left.\tilde{\Gamma}_{em}^0\,e^{-\beta\left(\omega_{f'f}-\omega_{cav}\right)}\frac{1}{\Xi}\left(\frac{\tilde{\Gamma}_{em}^0}{\Gamma_{loss}}e^{\beta\omega_{cav}}\right)^{N}e^{-\beta \omega_{f}}\right] =0,
\end{multline}
but it is crucial to keep in mind that this thermal-like distribution does not arises from any real thermalization process, but is a consequence of the specific form chosen for the pumping and dissipation. The application of this concept to the study of effective thermalization effects in a driven-dissipative non-Markovian condensate in the weakly interacting regime will be the subject of a future work, also with an eye to photon~\cite{Weitz} and polariton~\cite{BEC,Bajoni} Bose-Einstein condensation experiments.

\begin{figure}
\begin{center}
\includegraphics[width=0.25\columnwidth,clip]{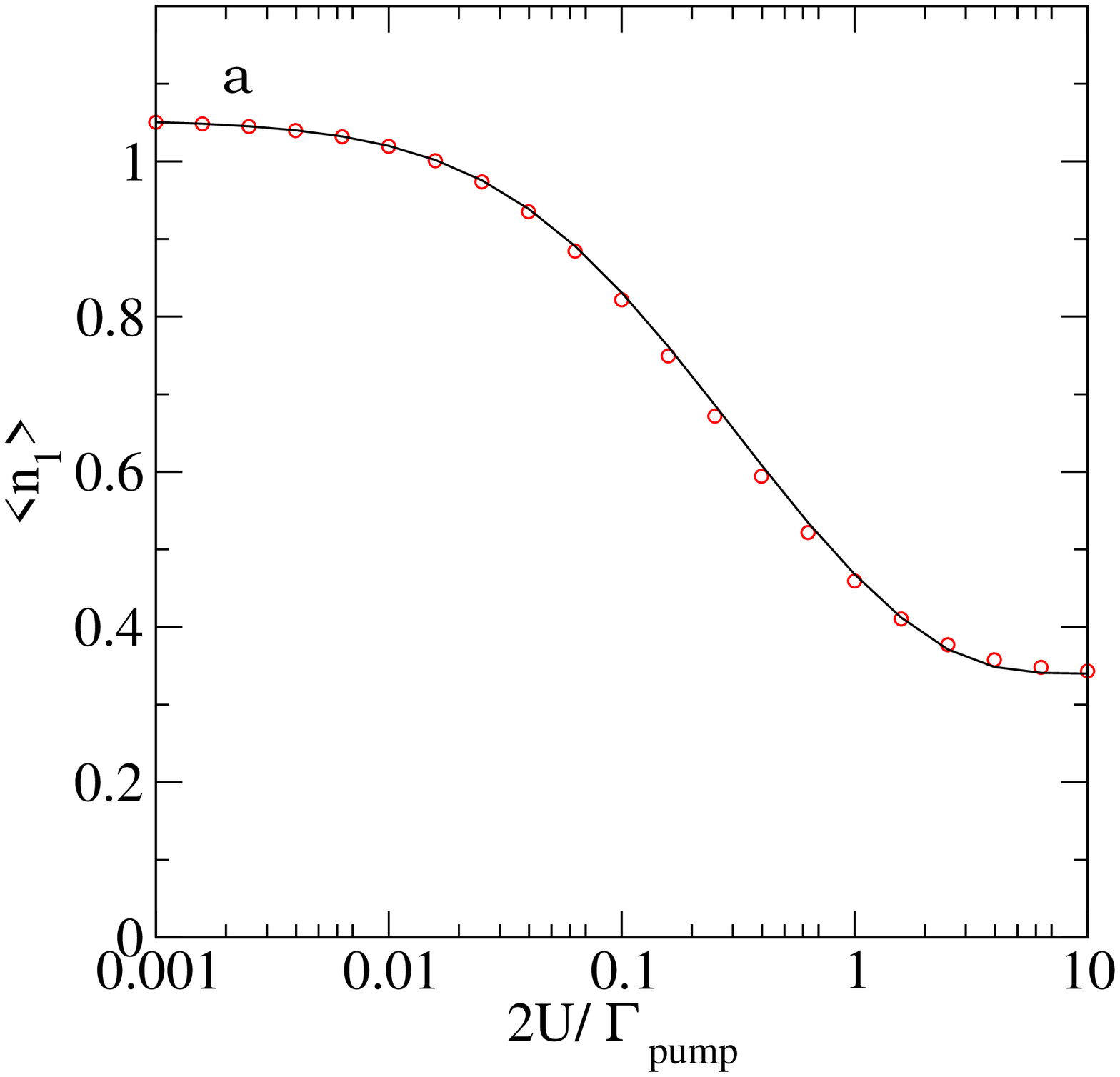}
\hspace{1cm}
\includegraphics[width=0.25\columnwidth,clip]{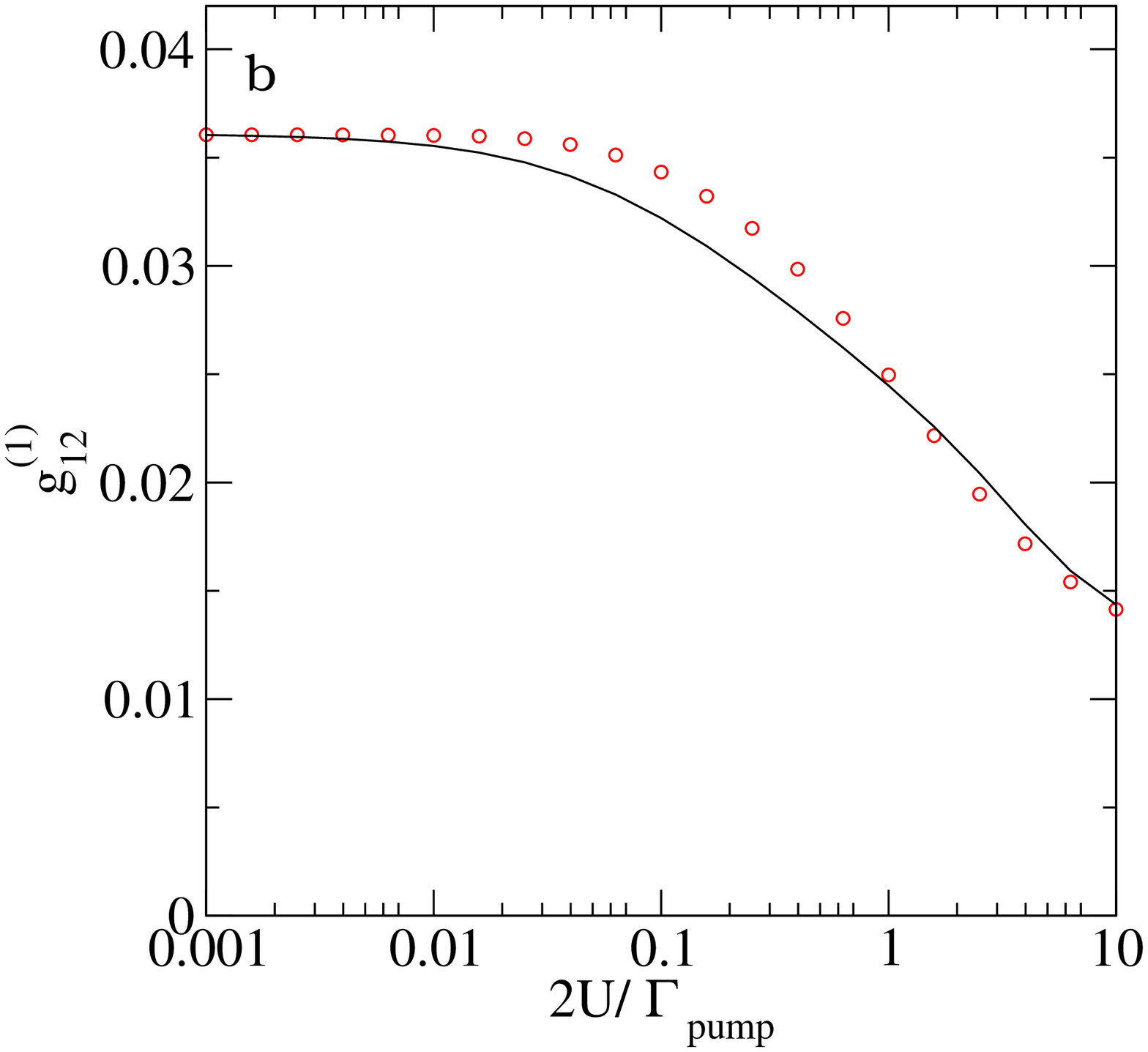}
\hspace{1cm}
\includegraphics[width=0.25\columnwidth,clip]{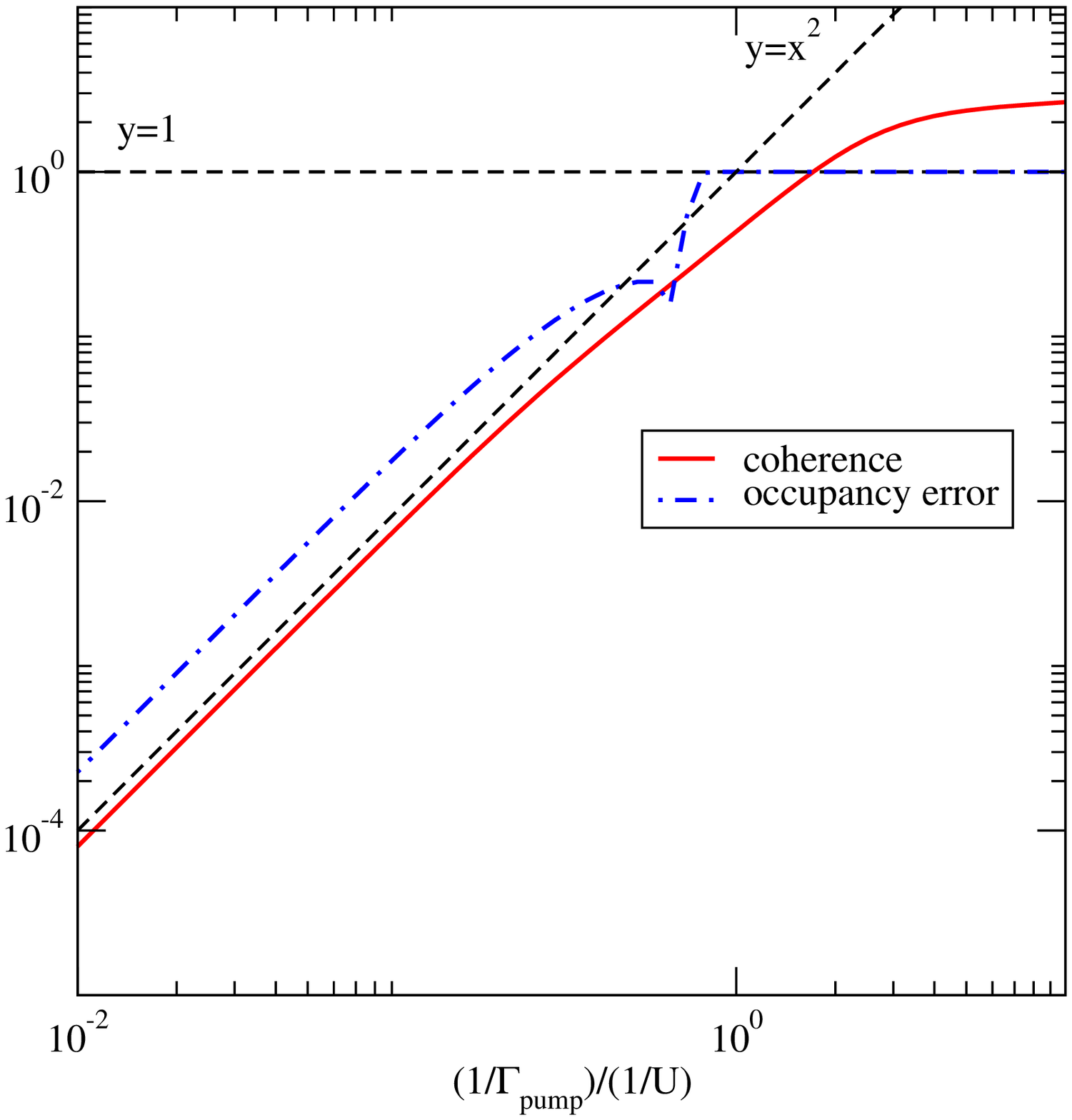}
\end{center}
\caption{
Left and center panels: average number of photons $n_1=\langle a_1^\dagger a_1\rangle$ (left) and spatial coherence $g_{1,2}^{(1)}= \langle a_1^\dagger a_2\rangle/\langle a_1^\dagger a_1\rangle$ (center) in a two cavity system with small $U/\Gamma_{pump}$ and $J/\Gamma_{pump}$ as a function of the non linearity $U$ at fixed $\Gamma_{pump}$. In red dots, exact resolution of the photonic master equation, and in black solid line the grand canonical ensemble ansatz. Parameters~: $2J/\Gamma_{pump}=0.02$, $2\Gamma_{loss}/\Gamma_{pump}=0.002$, $2\Gamma_{em}/\Gamma_{pump}=0.0014$, $2\delta/\Gamma_{pump}=0.6$. 
Right panel: purely photonic simulation of the relative quantum coherence between two arbitrarily chosen two-photon eigenstates $\rho_{ij}/\sqrt{\rho_{ii}\rho_{jj}}$ as a function of $1/\Gamma_{pump}$ (the result does not depend on the specific eigenstates considered). As expected, this coherence vanishes in $1/\Gamma^2_{pump}$ in the Markovian limit $1/\Gamma_{pump}\to 0$. The value above $1$ for large $1/\Gamma_{pump}$ signals breakdown of positivity of the density matrix as we move out of the validity regime of the purely photonic master equation.
Parameters: $J/\Gamma_{loss}=1$, $\Gamma_{em}/\Gamma_{loss}=0.5$, $\delta=-\Gamma_{loss}$, $U/\Gamma_{loss}=2$.
\label{fig:GCtest}}
\end{figure}
%J =0.01;
%G_loss=0.001;
%G_em=0.0007;
%P_ex=1;
%omega_ph=1;      
%omega_at=0.7;

A numerical test of this result for a two cavity system with a strong pumping $\Gamma_{pump}\gg U,J$ and a large enough photon number so to induce appreciable nonlinear effects is shown in Fig.\ref{fig:GCtest}. The results of this comparison are displayed in the left and central panels: excellent agreement between an exact resolution of the photonic master equation and the grand canonical ensemble ansatz is found in both the average photon number and the first-order coherence.

\subsection{Beyond the secular approximation}

In the weakly non-Markovian regime, the validity of the effective Grand-Canonical description can be extended outside the secular approximation according to the following arguments.
As a first step, we decompose the master equation as
\begin{equation}
\frac{d\rho}{dt}=[\mathcal{M}_0+\delta\mathcal{M}]\rho,
\end{equation}
where the super-operators $\mathcal{M}$ and $\delta\mathcal{M}$ act of the linear space of density matrices $\rho$ as
{\begin{equation}
\mathcal{M}_{0}[\rho]=-i\com H{\rho}+\frac{\Gamma_{loss}}{2}\sum_{i=1}^{k}\left[2a_{i}\rho a_{i}^{\dagger}-a_{i}^{\dagger}a_{i}\rho-\rho a_{i}^{\dagger}a_{i}\right] +\frac{\tilde{\Gamma}_{em}^0}{2}\sum_{i=1}^{k}\left[\hat{a}_{i}^{\dagger}\rho a_{i}+a_{i}^{\dagger}\rho\hat{a}_{i}-a_{i}\hat{a}_{i}^{\dagger}\rho-\rho\hat{a}_{i}a_{i}^{\dagger}\right],
\end{equation}}
and 
\begin{equation}
 \delta\mathcal{M}[\rho]=\frac{\tilde{\Gamma}_{em}^0}{2}\sum_{i=1}^{k}\left[\delta a_{i}^{\dagger}\rho a_{i}+a_{i}^{\dagger}\rho\delta a_{i}-a_{i}\delta a_{i}^{\dagger}\rho-\rho \delta a_{i} a_{i}^{\dagger}\right],
 \label{perturbation}
 \end{equation}
with 
\begin{equation}
\tilde{a}_{i}^{\dagger}=\frac{\tilde{\Gamma}_{em}^0}{\Gamma_{em}^0}\,\left( \hat{a}_{i}^{\dagger}+\delta a_{i}^{\dagger}\right),
\end{equation} and 
{\begin{equation}
\bra{f'}\hat{a}_{i}^{\dagger}\ket f=\left(e^{-\beta\Delta_{f'f}}-i\frac{\omega_{cav}-\omega_{at}}{\Gamma_{pump}}\right)\bra{f'}a_{i}^{\dagger}\ket f,
\end{equation}}
from which we deduce that 
{\begin{equation}
\bra{f'}\delta a_{i}^{\dagger}\ket f\underset{\Gamma_{pump}\to\infty}{=}\bra{f'}a_{i}^{\dagger}\ket f \, \left(-i\frac{\Delta_{f'f}}{\Gamma_{pump}}+\mathcal{O}\left(\frac{\Delta_{f',f}}{\Gamma_{pump}}\right)^{2}\right).
\label{deltacreation}
\end{equation} }
Using similar arguments to the Markovian case of \ref{app:Exact-stationary-solution}, we can easily show that the grand canonical distribution is a steady state of this modified $\mathcal{M}_{0}$ operator,
\begin{equation}
\mathcal{M}_{0}(e^{\beta N\mu}e^{-\beta H})=0,
\end{equation}
As the correction term $\delta\mathcal{M}$ vanishes in the Markovian limit proportionally to $1/\Gamma_{pump}$, we can calculate the lowest order correction to the steady state in $\delta\mathcal{M}$. Expanding the steady state in powers of $1/\Gamma_{pump}$ keeping a constant $(\omega_{cav}-\omega_{at})/\Gamma_{pump}$, we see easily that the first order corrections in eq. (\ref{deltacreation}) are purely imaginary so that populations are perturbed only to second order in {$\beta\Delta_{f'f}$}. In our Markovian limit, these corrections then vanish even if we perform simultaneously the Markovian and thermodynamic limit. 

Secondly, coherences (which are exactly zero in the Markovian case, see Sec.\ref{sec:Markov}) should be then proportional to {$1/\Gamma_{pump}$}. However, we have shown in \ref{app:quadratic coherences} that the linear contribution to coherences vanishes when we sum over all sites of the system. We conclude thus that in the weakly non Markovian limit, coherences between eigenstates of the hamiltonian are quadratic in $1/\Gamma_{pump}$ and therefore remain very small even out of the secular approximation. 

As a further verification of this analytical argument, in the right panel of Fig.\ref{fig:GCtest} we have shown the $\Gamma_{pump}$ dependence of the coherence between an arbitrary pair of two-photon states as well as the error in the population of an arbitrary eigenstate, between the true steady state and the grand canonical distribution. As expected on analytical grounds, both these quantities scale indeed as $\Gamma_{pump}^{-2}$.

From these arguments, we conclude that the breakdown of the secular approximation which occurs in the thermodynamic limit where the spectrum become continuous should not affect the effective thermalization of the steady state in the weakly non-Markovian regime of large $\Gamma_{pump}$. Even if the steady-state is not affected, we however expect that the relatively strong dissipation will significantly affect the the system dynamics. A complete study of this physics will be the subject of a future work. 

\section{Two cavities with strong non linearity}

\subsection{Towards Mott-insulator physics}

As a final example of application of our concepts, in this last section we present some preliminary results on the most interesting case of two strongly nonlinear cavities with $U\gg\Gamma_{pump}$: extending the photon-number selectivity idea to the many-cavity case, we look for many-body states that resemble a Mott insulator~\cite{Bloch2008,RMP,Hartmann_rev}. As in the single cavity case, the strong pumping $\Gamma_{em}\gg \Gamma_{loss}$ would favour a large occupations of sites, but is counteracted by the effect of the nonlinearity $U\gg \Gamma_{pump}$ which {sets} an upper bound to the occupation: the result is a steady-state with a well-defined number of photons per cavity.

\begin{figure}
\begin{center}
\includegraphics[scale=0.3,clip]{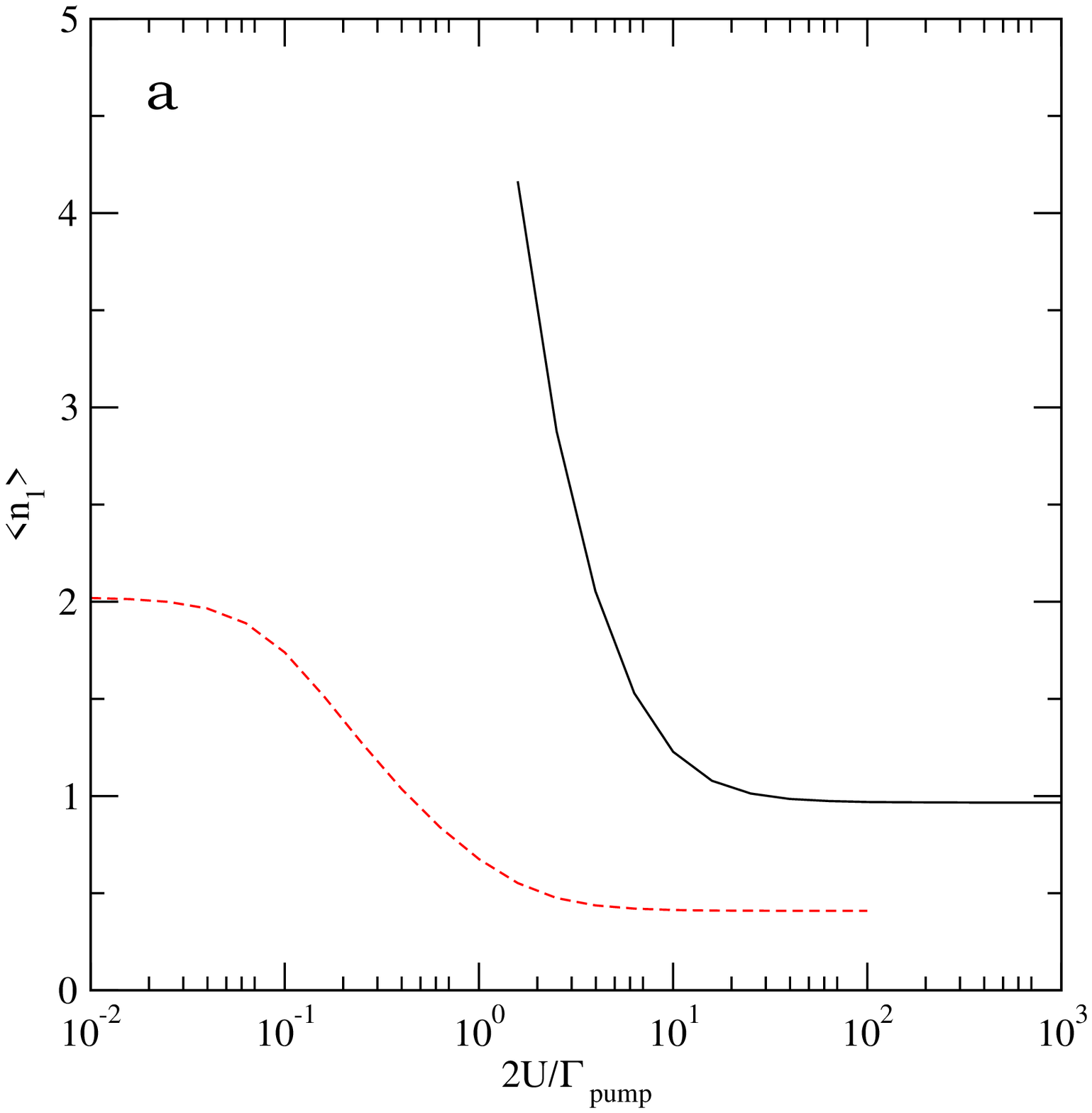} \hspace{1cm}
\includegraphics[scale=0.3,clip]{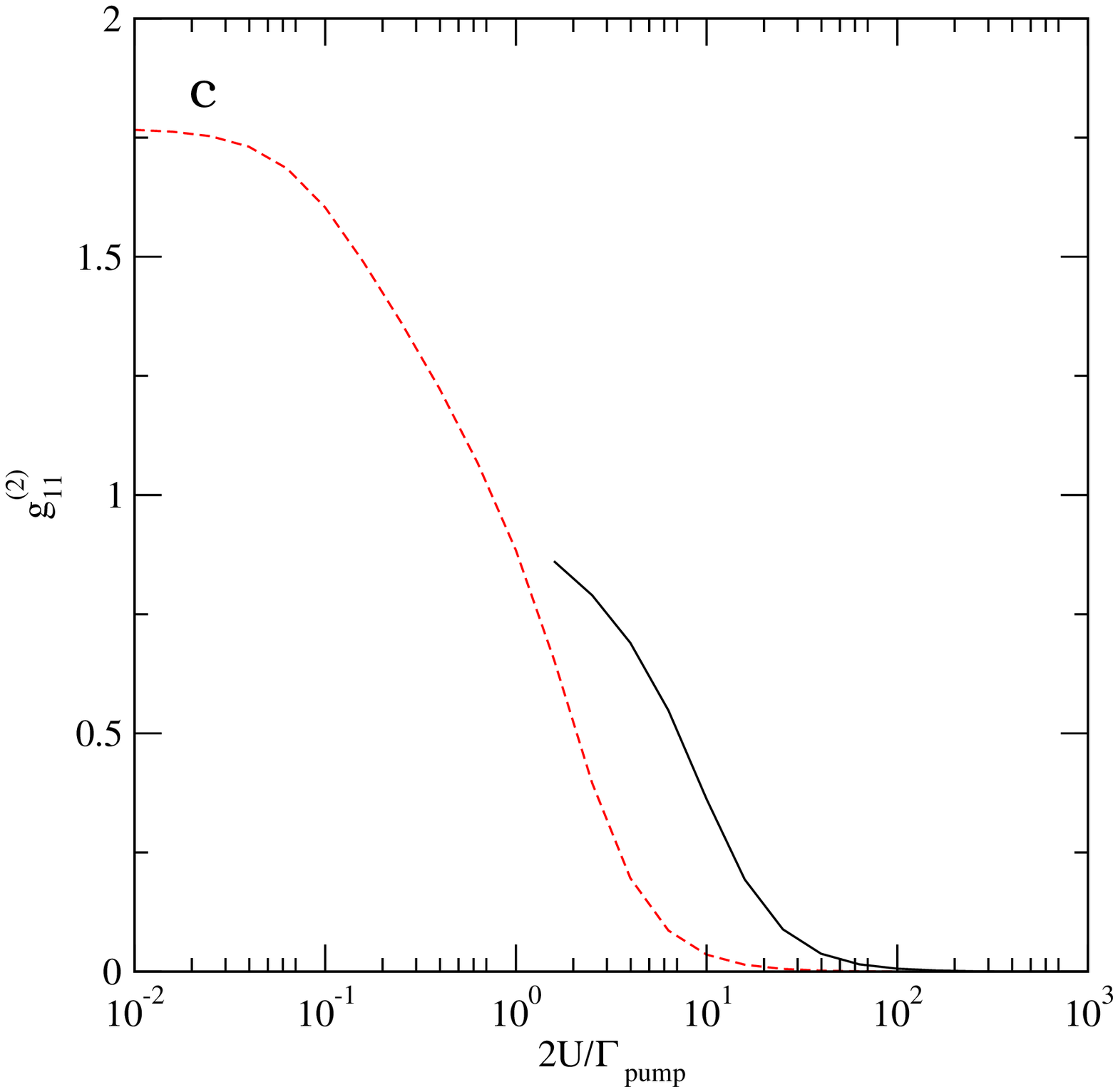} \\
\includegraphics[scale=0.3,clip]{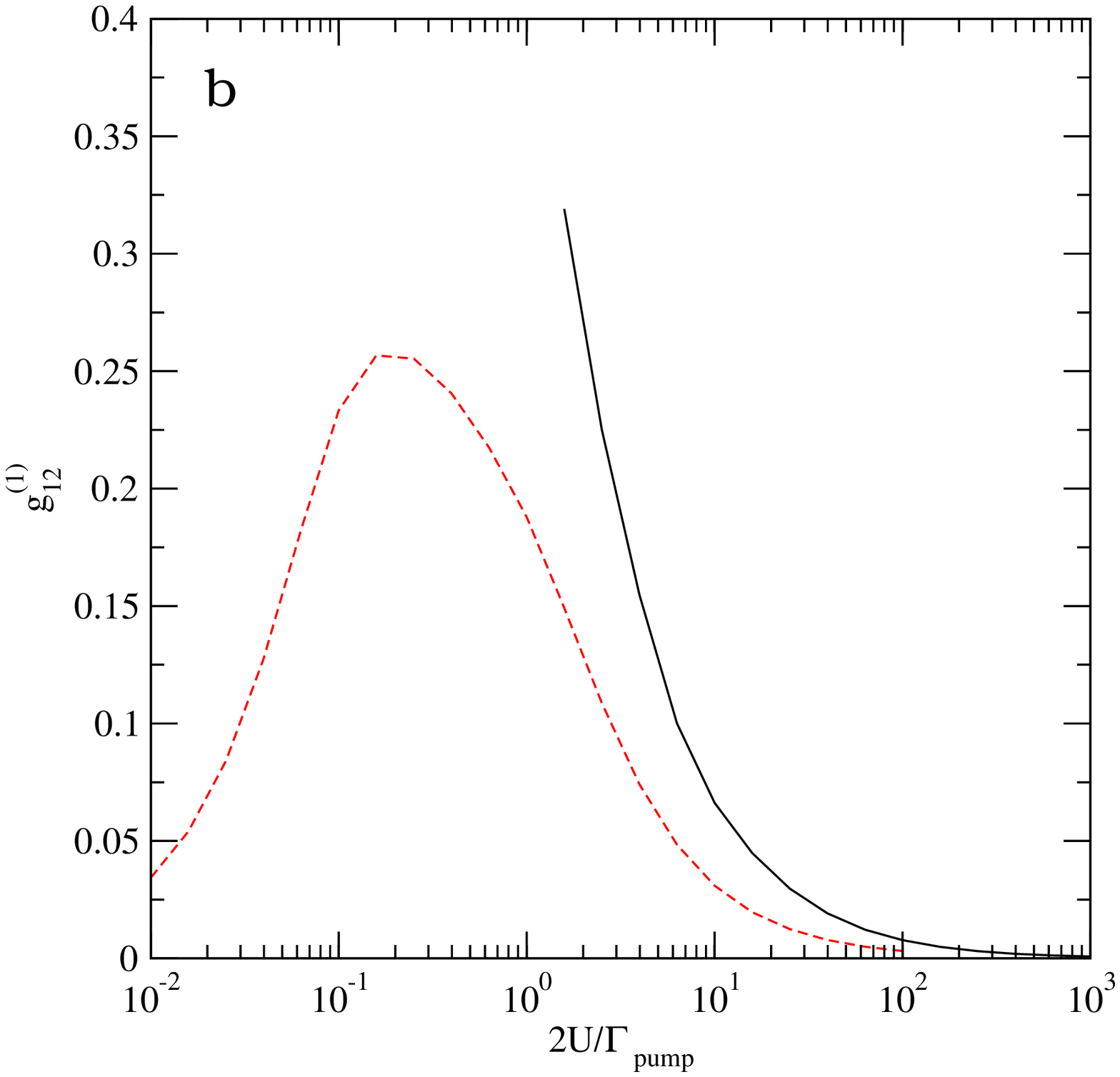} \hspace{1cm}
\includegraphics[scale=0.3,clip]{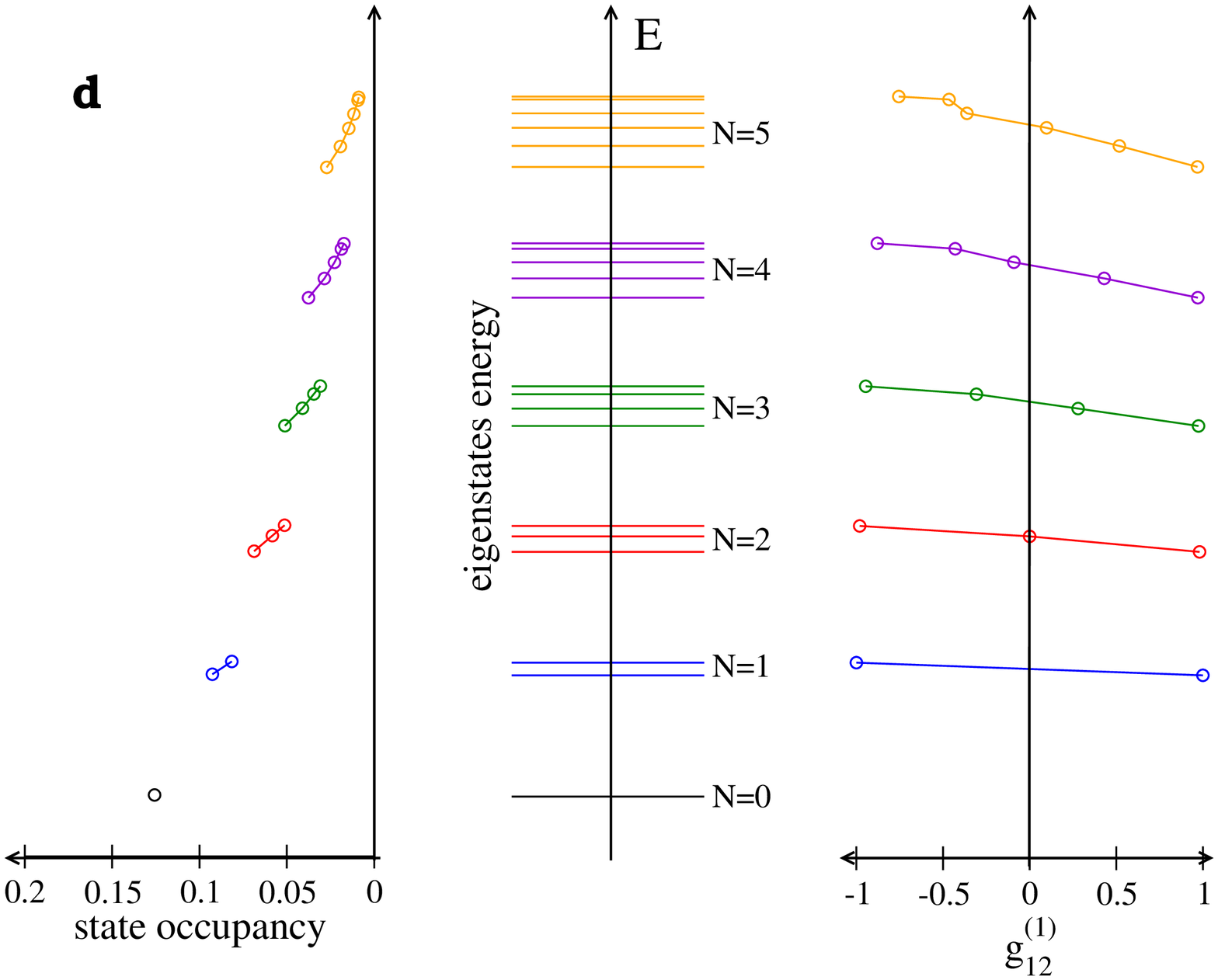}
 \end{center}
\caption{Purely photonic simulations of steady-state observables as a function of $2U/\Gamma_{pump}$ in a two-cavity system: (a) average number of photons $n_{1}=\langle{a_{1}^{\dagger}a_{1}}\rangle$, (b) one-site two-body correlation function $g_{1,1}^{(2)}=\langle{a_{1}^{\dagger}a_{1}^{\dagger}a_{1}a_{1}}\rangle=\langle{n_{1}(n_{1}-1)}\rangle$, (c) inter-site one-body correlation function $g_{1,2}^{(1)}={\langle a_{1}^{\dagger}a_{2}\rangle }/{\langle a_{1}^{\dagger}a_{1}\rangle }$. Parameters: $2J/\Gamma_{pump}=0.2$, $2\Gamma_{loss}/\Gamma_{pump}=0.002$, $2\Gamma_{em}/\Gamma_{pump}=0.06$ (solid black line). Red dashed line, same simulation with a weaker $2\Gamma_{em}/\Gamma_{pump}=0.00144$. Panel (d), from left to right~:  state occupancy, energy and two site spatial coherence of the different eigenstates of the hamiltonian, at the maximum coherence point $2U/\Gamma_{pump}=0.16$ of the red dashed line.
% $\mathrm{log}_{10}(2U/\Gamma_{pump})=-0.8$.
%wat = wcav.
\label{fig:mott}}
%a b c
%black 
%2J/Gpump=0.2
%2Gcav/Gpump=0.002
%2Gem/Gpump=0.06
%wat = wcav
%red
%J =0.1;
%G_loss=0.001;
%G_em=0.00072;
%P_ex=1;
%omega_ph=1;      
%omega_at=1;      
%d
%log10(2U/\Gamma_{pump}) =-0.8;               
%J =0.1;
%G_loss=0.0010;
%G_em=0.00072;
%P_ex=1;
%omega_ph=20;      
%omega_at=20;     
%g12=0.307
\end{figure}

The result of numerical calculations based on the photonic master equation are shown as black lines in Fig.\ref{fig:mott}(a-c) in the $\omega_{cav}=\omega_{at}$ case: for a high emission rate $\Gamma_{em}^0$ and a strong non linearity $U$, signatures of the desired Mott state with one particle per site are visible in the steady-state average number of photons  that tends to $1$ for a strong nonlinearity $U$ [panel (a)], in the probability of double occupancy  that tends to $0$ [panel (b)], and in the one-body coherence between the two sites that also tends to $0$ [panel (c)]. 

While these results are a strong evidence of $N_0=1$ Mott state, a similar calculation for larger $N_0 \geq 2$ Mott states is made much more difficult by metastability issues and the Mott state would typically have a finite lifetime. As in the single cavity case, we expect that this problem could be fixed by adding several atomic species on resonance with the different photonic transitions below $N_0$.

Based on this preliminary analysis, we can attempt to make some claims on the structure of the non-equilibrium phase diagram of our model. As for $J=0$ one can efficiently create a Fock state in each cavity, we expect that for small $J$ the system will remain in a sort of Mott state. On the other hand, in the weakly interacting regime we expect the system to display a coherent Bose-Einstein condensate~\cite{PRL10}. In between, one can anticipate that system should display some form of non-equilibrium Mott-Superfluid transition. Analytical and numerical studies in this direction are in progress.

\subsection{An unexpected mechanism for coherence}

The red dashed lines in the same panels Fig.\ref{fig:mott}(a-c) show the same simulation for a weaker emission rate $\Gamma_{em}^0$, which allows to consider weaker values of the nonlinearity without increasing too much the photon number. In particular, in panel (c) we see that the non-negligible value of $2J/\Gamma_{pump}$ is responsible for a  significant spatial coherence between the two sites, which attains a maximum value $g^{(1)}_{12}\approx 0.26$ for an interaction strength $2U/\Gamma_{pump}\simeq 0.16$ of the same order of magnitude as the tunnel coupling $2J/\Gamma_{pump}=0.2$. 

The quite unexpected appearance of this coherence can be understood as follows. On one hand, in the absence of tunneling $J=0$, all the dynamics is local and we do not expect any spatial coherence. On the other hand, in the absence of interactions $U=0$ and for zero detuning, symmetric and anti-symmetric states are equally close to resonance (albeit with opposite detuning) and then equally populated, so there should not be any coherence either. However in presence of both tunnelling and small interactions (i.e. for $J,U\neq 0$ and $U\ll J$), the energy of all eigenstates (symmetric/anti-symmetric states with various photon numbers) is perturbatively shifted in the upward direction by (small) interactions $U$. As a result, symmetric states, which are below the resonance, get closer to resonance and become more populated than the anti-symmetric ones, which get farther to the resonance and are thus depleted. As one can see in the plot of the energy, the spatial coherence and the steady-state occupancy of the different eigenstates shown in Fig.\ref{fig:mott}(d) for the maximum coherence point, this induces an overall positive coherence between the two sites.

Even though the nonlinearity is only active for states with at least two photons, it is interesting to note that also in the $N=1$ manifold the antisymmetric state is less populated than the symmetric one. This population unbalance is inherited from the one in the above-lying $N>1$ states, as the decay preferentially occurs into the symmetric state. Since no coherence is expected in both limiting cases of purely interacting $U\gg J$ and non interacting $U/0=0$ photons, the maximum of the coherence is obtained when interactions and tunelling are of the same magnitude, $U\approx J$: this result is clearly visible in panel Fig.\ref{fig:mott}(c).

Investigation of this many-body physics in the more interesting case of larger arrays which can accommodate a larger number of photons requires sophisticated numerical techniques to deal with the dynamics in a huge Hilbert space~\cite{Pisa,corner} and will be the subject of future work. A very exciting advance in this direction was recently published in~\cite{Kapit} for strongly interacting photons in the presence of a synthetic gauge field for light: analogously to the Mott insulator state studied here, the combination of the effectively frequency-dependent pumping (obtained via a two-photon pumping in the presence of an auxiliary lattice) and the many-body energy gap was predicted to generate and stabilize fractional quantum Hall states of light.

\section{Conclusions}
\label{sec:Conclu}

In this work we have proposed and characterized a novel scheme to generate strongly correlated states of light in strongly nonlinear cavity arrays. Photons are incoherently injected in the cavities using population-inverted two-level atoms, which preferentially emit photons around their resonance frequency. The resulting frequency-dependence of the gain will be the key element to generate and stabilize the desired quantum state. A manageable theoretical description of the system is obtained using projective methods, which allow to eliminate the atomic degrees of freedom and describe the non-Markovian photonic dynamics in terms a generalized master equation. 

The efficiency of the our pumping scheme to generate specific quantum states is first validated on a single-cavity system: for weak nonlinearities, a novel mechanism for optical bistability is found. For strong nonlinearities, Fock states with a well-defined photon number can be generated with small number fluctuations. 

In the general many cavity case, in the weakly non-Markovian case the steady-state of the system recovers a Grand-Canonical distribution with an effective chemical potential determined by the pumping strength and an effective inverse temperature proportional to the non-Markovianity: This very general results may have application to explain apparent thermalization in recent photon and polariton condensation experiments. 

Finally, the power of a frequency-dependent pumping to generate strongly correlated states of light is illustrated in the case of a strongly nonlinear two-cavity system which, in the strongly non-Markovian regime, can be driven into a state that closely reminds a Mott-insulator state. A general study of the potential and of the limitations of the frequency-dependent gain to generate generic strongly correlated states with many photons will be the subject of future work.

\section{Acknowledgments}
IC acknowledges financial support by the ERC through the QGBE grant, by the EU-FET Proactive grant AQuS (Project No.640800), and by the Provincia Autonoma di Trento, partly through the project ``On silicon chip quantum optics for quantum computing and secure communications'' (``SiQuro''). Continuous discussions with Alessio Chiocchetta, Hannah Price, Alberto Amo, Jacqueline Bloch, and Mohammad Hafezi are warmly acknowledged.

\appendix

\appendix

\section{Derivation of the purely photonic master equation via projective methods}
\label{app:projective}
In this Appendix, we give more details on the derivation of the photonic master equation (\ref{eq:photon_only}). Starting from the full atom-cavity master equation (\ref{eq:evinitio}), we show how  for a sufficiently small atom-cavity coupling $\Omega_R$ the atomic degrees of freedom can be eliminated. The frequency-dependence of the atomic amplification is then accounted for as a modified Lindblad term (\ref{eq:gainmarkov}). Our treatment is based on the discussion in the textbook~\cite{Breuer}.

\subsection{\emph{General formalism}}
\label{app:proj_gen}
We consider a quantum system which undergoes dissipative processes. As it is not isolated, its state can not be described by a wave function but by a density matrix $\rho$ evolving according to the master equation~:
\begin{equation}
\partial_{t}\rho=\mathcal{L}(\rho(t)),
\end{equation}
where $\mathcal{L}$ is some linear ``super-operator'' acting on the space of density matrices. Given an arbitrary initial density matrix $\rho(t_0)$, the density matrix $\rho$ at generic time $t$ 
is equal to $\rho(t)=e^{\mathcal{L}(t-t_{0})}\rho(t_{0})$.\\

Now we are only interested in some part of the density matrix, which
can represent some subsystem. This can be described by a projection
operation on the density matrix $\mathcal{P}\rho$ . We call $\mathcal{Q}=1-\mathcal{P}$
the complementary projector. We decompose the Lindblad operator $\mathcal{L}$ in two parts $\mathcal{L}_{0}$ and $\delta\mathcal{L}$ such that:
\begin{equation}
\left\{ \begin{array}{l}
\mathcal{\mathcal{L}=\mathcal{L}}_{0}+\delta\mathcal{L}\\
\mathcal{P}\mathcal{L}_{0}\mathcal{Q}=\mathcal{Q}\mathcal{L}_{0}\mathcal{P}=0\\
\mathcal{P}\,\delta\mathcal{L}\,\mathcal{P}=0.
\end{array}\right.\label{eq:condition projector}
\end{equation}
Such a decomposition is always possible. 

Then we define a generalised interaction picture for the density matrix and for generic superoperators $\mathcal{A}$ with respect to the evolution described by the free $\mathcal{L}_{0}$ and the initial time $t_{0}$: 
\begin{equation}
\left\{ \begin{array}{l}
\hat{\rho}(t)=e^{-\mathcal{L}_{0}(t-t_{0})}\rho(t)\\
\hat{\mathcal{\mathcal{A}}}(t)=e^{-\mathcal{L}_{0}(t-t_{0})}\mathcal{A}e^{\mathcal{L}_{0}(t-t_{0})}.
\end{array}\right.
\end{equation}

As discussed in \cite{Breuer}, we can get an exact closed master
equation for the projected density matrix in the interaction picture
\begin{equation}
\partial_{t}\mathcal{P}\hat{\rho}(t)=\int_{t_{0}}^{t}dt'\Sigma(t,t')\mathcal{P}\hat{\rho}(t'), \label{eq: ev self int}
\end{equation}
which translates to 
\begin{eqnarray}
\partial_{t}\mathcal{P}\rho(t) & = & \mathcal{L}_{0}(\rho(t))+\int_{t_{0}}^{t}dt'\tilde{\Sigma}(t-t')\mathcal{P}\rho(t')\label{eq:ev self schro-1}
\end{eqnarray}
in the Schrodinger picture. In the interaction picture, the self energy operator $\Sigma$ is defined as:
\begin{equation}
\Sigma(t,t') = \sum_{n=2}^{\infty}\int_{t'}^{t}\int_{t'}^{t_{1}}..\int_{t'}^{t_{n-1}}dt_{1}..dt_{n}\mathcal{P}\delta\hat{\mathcal{L}}(t)\mathcal{Q}\delta\hat{\mathcal{L}}(t_{1})\mathcal{Q}\delta\hat{\mathcal{L}}(t_{2})...\mathcal{Q}\delta\hat{\mathcal{L}}(t_{n})\mathcal{Q}\delta\hat{\mathcal{L}}(t')\mathcal{P}
\end{equation}
and results from the coherent sum over the processes leaving from $\mathcal{P}$,
remaining in $\mathcal{Q}$ and then coming back finally to $\mathcal{P}$.
In the Schrodinger representation, we have :
\begin{equation}
\tilde{\Sigma}(t-t')=e^{\mathcal{L}_{0}(t-t_{0})}\Sigma(t,t')e^{-\mathcal{L}_{0}(t'-t_{0})}=\Sigma(0,t'-t)e^{\mathcal{L}_{0}(t-t')}.\label{app:schrodinger-self-energy}
\end{equation}

We call $\tau_{c}=1/\Delta\omega$ the characteristic decay time /
inverse linewidth for the self energy, which corresponds in general to
the correlation time of the bath, and we estimate the rate of dissipative processes as $\Gamma\simeq\Sigma\tau_{c}=\int_{t_{0}}^{\infty}dt\Sigma(t,t_{0})$. We
put ourselves in the regimes in which, with respect to these dissipative
processes, the bath has a short memory, ie $\Gamma\ll\Delta\omega$.
In that regime the density matrix in the interaction picture is almost
constant over that time $\tau_{c}$. Furthemore, if $t-t_{0}\gg\tau_{c}$
then the integral in eq (\ref{eq: ev self int}) can be extended from $-\infty$
to $t$. From this equation and from (\ref{eq: ev self int}), we get an
equation of evolution for the density matrix which is local in time~:
\begin{eqnarray}
\partial_{t}\mathcal{P}\hat{\rho}(t) & = & \int_{0}^{\infty}d\tau\Sigma(t,t-\tau)\mathcal{P}\hat{\rho}(t)\nonumber\\
 & = & \int_{0}^{\infty}d\tau\left[e^{-\mathcal{L}(t-t_{0})}\Sigma(0,-\tau)e^{\mathcal{L}(t-t_{0})}\right]\mathcal{P}\hat{\rho}(t)\nonumber\\
 & = & e^{-\mathcal{L}(t-t_{0})}\int_{0}^{\infty}d\tau\Sigma(0,-\tau)\mathcal{P}\rho(t).
\end{eqnarray}
In the Schrodinger picture this gives the time-local master equation~:
\begin{equation}
\partial_{t}\mathcal{P}\hat{\rho}(t) = \left[\mathcal{L}_{0}+\int_{0}^{\infty}d\tau\Sigma(0,-\tau)\right]\,\mathcal{P}\rho(t)=
\mathcal{L}_{eff}\mathcal{P}\rho(t),\label{eq:approx eff ev}
\end{equation}
with
\begin{equation}
\mathcal{L}_{eff}=\mathcal{L}_{0}+\int_{0}^{\infty}d\tau\Sigma(0,-\tau).
\end{equation}
It is worth stressing that while the bath is Markovian with respect to dissipative processes induced by the perturbation $\int_{0}^{\infty}d\tau\Sigma(0,-\tau)$ , no Markovian approximation has been made with respect to the dynamics due to $\mathcal{L}_{0}$, which can still be fast. For the specific system under consideration in this work, this means that the emission rate $\Gamma_{em}$ has to be slow with respect to the gain bandwidth set by the atomic pumping rate $\Gamma_{pump}$, which is the case in the weak coupling limit $\sqrt{N_{at}}\Omega_{R}\ll\Gamma_{pump}$, but no restriction is to be imposed on the parameters $U$, $J$ and $\omega_{cav}-\omega_{at}$ of the Hamiltonian, which can be arbitrarily large. This means that the physics can be strongly non-markovian with respect to the Hamiltonian photonic dynamics.

\subsection{\emph{Application to the array of cavities} }
\subsubsection*{Preliminary calculations}

With the notation from section \ref{sec:Presentation-of-the}, we choose
the projectors in the form~:
\begin{equation}
\mathcal{P}\rho=\ket{e_{1}^{(1)}e_{2}^{(1)}e_{3}^{(1)}...}\bra{e_{1}^{(1)}e_{2}^{(1)}e_{3}^{(1)}...}\otimes Tr_{at}(\rho),
\end{equation}
where we have performed a partial trace over the atoms, and then make the tensor
product of the density matrix and the atomic density matrix with all
atoms in the excited state. We chose this particular projector because in the weak atom-cavity coupling regime, we expect atoms to be repumped almost immediately after having emitted a photon in the cavity array, and thus to be most of the time in the excited state. Moreover this projection operation gives us direct access to the photonic density matrix, and thus we do not lose any information on photonic statistics. With the notation of the previous section we have~:
\begin{equation}
 \mathcal{L}(\rho)=-i\com{H_{ph}+H_{at}+H_{I}}{\rho}+\mathcal{L}_{diss}(\rho),
 \end{equation}
with 
\begin{equation}
\mathcal{L}_{diss}=\mathcal{L}_{pump, at}+\mathcal{L}_{loss, cav} .
\end{equation}
We decompose $\mathcal{L}$ in two contributions. 
The first one is~:
\begin{equation}
\mathcal{L}_{0}(\rho) =  -i\com{H_{ph}+H_{at}}{\rho}+\mathcal{L}_{loss, cav}(\rho) -\mathcal{A}(\rho)+\mathcal{P}\mathcal{A}\mathcal{Q}(\rho)\label{app:L0}
\end{equation}
with 
\begin{equation}
\mathcal{A}(\rho)=\frac{\Gamma_{pump}}{2}\sum_{i=1}^{k}\sum_{l=1}^{N_{at}}\left[\sigma_{i}^{-(l)}\sigma_{i}^{+(l)}\rho+\rho\sigma_{i}^{-(l)}\sigma_{i}^{+(l)}\right].
\end{equation}
The superoperator $\mathcal{L}_{0}$ verifies the condition (\ref{eq:condition projector}):
The last term in the expression of eq.(\ref{app:L0}) comes from the fact that the pumping term $\mathcal{A}$ in $\mathcal{L}_0$ does not verify this condition: as a result, we have to remove the part unfixed by projector and put it in the other operator
:
\begin{eqnarray}
\delta\mathcal{L}(\rho) & = & -i\com{H_{I}}{\rho}+\frac{\Gamma_{pump}}{2}\sum_{i=1}^{k}\sum_{l=1}^{N_{at}}2\sigma_{i}^{+(l)}\rho\sigma_{i}^{-(l)}-\mathcal{P}\mathcal{A}\mathcal{Q}(\rho).
\end{eqnarray}

These two operators then satisfy to the conditons (\ref{eq:condition projector}), and we can apply the projection method to get the evolution of $\mathcal{P}\rho(t)$, that is of $Tr_{at}(\rho)(t)$.
As we are interested in the regime in which $\Gamma_{pump}\gg\sqrt{N_{at}}\Omega_{R},\,\Gamma_{loss}$, we will compute the self energy at the lowest non zero order of these two latter parameters. Since $\Gamma_{loss}$ quantifies the photonic loss rate, we will approximate the photonic dynamics as being a Hamiltonian one during the time while the atom is reinjected in the excited state, ie during the characteristic time $1/\Gamma_{pump}$ of the integration kernel of eq.(\ref{eq:ev self schro-1}). To this order of precision, the calculation for one cavity is easily generalizable to $k$ cavities, thus we will restrict for simplicity to the case of a single cavity containing a single two-level atom, $N_{at}=1$.
\subsubsection*{Self energy calculation~:}

We are going to calculate the self energy to the lowest order in $\Omega_{R}$. We have 
\begin{equation}
\delta\mathcal{L}=\mathcal{L}_{pump}-i(H^{+}+H^{-})_{L}+i(H^{+}+H^{-})_{R}-\mathcal{P}\mathcal{A}\mathcal{Q},
\end{equation}
with 
\begin{equation}
\left\{ \begin{array}{l}
\mathcal{L}_{pump}(\rho)=\Gamma_{pump}\sigma^{+}\rho\sigma^{-}\\
H^{+}=\Omega_{R}\sigma^{+}a\\
H^{-}=\Omega_{R}\sigma^{-}a^{\dagger}\\
\end{array}\right.
\end{equation}
By $(H^{\pm})_{L/R}$ we intend the superoperator multiplying a matrix $\rho$ by the matrix $H^{\pm}$ on its left/right.
First we have $\mathcal{L}_{pump}\mathcal{P}=\mathcal{P}\mathcal{A}\mathcal{Q}\mathcal{P}=H_{L}^{+}\mathcal{P}=H_{R}^{-}\mathcal{P}=0$,
so starting from a projected state $\mathcal{P}\rho$, we have to start with $H_{L}^{-}$
or $H_{R}^{+}$. 
In fact to the lowest order in $\Omega_{R}$ the non zero contributions
to the self energy are~:

\begin{equation}
\begin{array}{l}
A=-\mathcal{P}H_{L}^{+}H_{L}^{-}(t'-t)\mathcal{P}\\
B=-\mathcal{P}H_{R}^{-}H_{R}^{+}(t'-t)\mathcal{P}\\
C=\mathcal{P}H_{R}^{+}H_{L}^{-}(t'-t)\mathcal{P}\\
D=\mathcal{P}H_{L}^{-}H_{R}^{+}(t'-t)\mathcal{P}\\
E=\int_{t_{'}}^{t}d\tilde{t}\,\mathcal{P}\mathcal{L}_{pump}(t)\mathcal{Q}H_{R}^{+}(\tilde{t}-t)H_{L}^{-}(t'-t)\mathcal{P}\\
F=\int_{t'}^{t}d\tilde{t}\,\mathcal{P}\mathcal{L}_{pump}(t)\mathcal{Q}H_{L}^{+}(\tilde{t}-t)H_{R}^{-}(t'-t)\mathcal{P}\\
G=-\int_{t_{'}}^{t}d\tilde{t}\,\mathcal{P}\mathcal{A}\mathcal{Q}H_{R}^{+}(\tilde{t}-t)H_{L}^{-}(t'-t)\mathcal{P}\\
H=-\int_{t_{'}}^{t}d\tilde{t}\,\mathcal{P}\mathcal{A}\mathcal{Q}H_{L}^{-}(\tilde{t}-t)H_{R}^{+}(t'-t)\mathcal{P},
\end{array}
\end{equation}

with 
\begin{equation}
\Sigma(0,t'-t)=A+B+C+D+E+F+G+H.
\end{equation}

We then calculate the different processes, applied on some projected matrix $\mathcal{P}\rho$:

\begin{eqnarray}
A(\mathcal{P}\rho) & = & -\Omega_{R}^{2}e^{(i\omega_{at}-\Gamma_{pump}/2)(t-t')}a a^{\dagger}(t'-t)\mathcal{\mathcal{P}}\rho\nonumber\\
B(\mathcal{P}\rho) & = & -\Omega_{R}^{2}e^{-(i\omega_{at}+\Gamma_{pump}/2)(t-t')}\mathcal{\mathcal{P}}\rho a(t'-t)a^{\dagger}\nonumber\\
C(\mathcal{P}\rho) & = & \Omega_{R}^{2}e^{(i\omega_{at}-\Gamma_{pump}/2)(t-t')}a^{\dagger}(t'-t)\mathcal{P}\rho a\nonumber\\
D(\mathcal{P}\rho) & = & \Omega_{R}^{2}e^{(-i\omega_{at}+\Gamma_{pump}/2)(t-t')}a^{\dagger}\mathcal{P}\rho a(t'-t)\nonumber\\
E(\mathcal{P}\rho) & = & \Gamma_{pump}\Omega_{R}^{2}\int_{t_{'}}^{t}d\tilde{t}\,e^{(-i\omega_{at}-\Gamma_{pump}/2)(t-\tilde{t})}\nonumber\\
&&e^{(i\omega_{at}-\Gamma_{pump}/2)(t-t')}a^{\dagger}(t'-t)\mathcal{P}\rho a(\tilde{t}-t)\nonumber\\
F(\mathcal{P}\rho) & = & \Gamma_{pump}\Omega_{R}^{2}\int_{t_{'}}^{t}d\tilde{t}\,e^{(i\omega_{at}-\Gamma_{pump}/2)(t-\tilde{t})}\nonumber\\
&&e^{(-i\omega_{at}-\Gamma_{pump}/2)(t-t')}a^{\dagger}(\tilde{t}-t)\mathcal{P}\rho a(t'-t)\nonumber\\
G(\mathcal{P}\rho) & = & - \Gamma_{pump}\Omega_{R}^{2}\int_{t_{'}}^{t}d\tilde{t}\,e^{(-i\omega_{at}-\Gamma_{pump}/2)(t-\tilde{t})}\nonumber\\
&&e^{(i\omega_{at}-\Gamma_{pump}/2)(t-t')}a^{\dagger}(t'-t)\mathcal{P}\rho a(\tilde{t}-t)=-E(\mathcal{P}\rho)\nonumber\\
H(\mathcal{P}\rho) & = & - \Gamma_{pump}\Omega_{R}^{2}\int_{t_{'}}^{t}d\tilde{t}\,e^{(i\omega_{at}-\Gamma_{pump}/2)(t-\tilde{t})}\nonumber\\
&&e^{(-i\omega_{at}-\Gamma_{pump}/2)(t-t')}a^{\dagger}(\tilde{t}-t)\mathcal{P}\rho a(t'-t)=-F(\mathcal{P}\rho)\nonumber\\
\end{eqnarray}
where by $ a(t'-t)$ we intend the evolution of the photonic annihlation operator in the photonic hamiltonian interaction picture (we remind that we neglected photonic losses during the integration time). We see that the last four contribution cancel each other, and that only the first four contributions remain.
\subsubsection*{Master equation}
Using the expression for the self-energy $\Sigma(t)$ derived in the last section, as well as general results on the master equation obtained by projective methods in Sec.\ref{app:proj_gen}, we then obtain the (temporally non-local) master equation~:
% \begin{eqnarray}
% \partial_{t}\mathcal{P}\rho & = & -i\left[H_{ph},\mathcal{P}\rho\right]+\mathcal{L}_{\Gamma}(\mathcal{P}\rho)\\
%  &  & +\Omega_{R}^{2}\int_{0}^{\infty}d\tau e^{(i\omega_{at}-\Gamma_{pump}/2)\tau}\nonumber\\
% &&\phantom{ -\Omega_{R}^{2}\int_{0}^{\infty}d\tau}a^{\dagger}\left(e^{\mathcal{L}_{0}(\tau)}\mathcal{\mathcal{P}}\rho(t-\tau)\right)a(-\tau)\nonumber\\
%  &  & +\Omega_{R}^{2}\int_{0}^{\infty}d\tau e^{-(i\omega_{at}+\Gamma_{pump}/2)\tau}\nonumber\\
%  &&\phantom{ -\Omega_{R}^{2}\int_{0}^{\infty}d\tau}a^{\dagger}(-\tau)\left(e^{\mathcal{L}_{0}(\tau)}\mathcal{\mathcal{P}}\rho(t-\tau)\right)a\nonumber\\
%  &  & -\Omega_{R}^{2}\int_{0}^{\infty}d\tau e^{(i\omega_{at}-\Gamma_{pump}/2)\tau}\nonumber\\
%  && \phantom{ -\Omega_{R}^{2}\int_{0}^{\infty}d\tau}aa^{\dagger}(-\tau)\left(e^{\mathcal{L}_{0}(\tau)}\mathcal{\mathcal{P}}\rho(t-\tau)\right)\nonumber\\
%  &  & -\Omega_{R}^{2}\int_{0}^{\infty}d\tau e^{-(i\omega_{at}+\Gamma_{pump}/2)\tau}\nonumber\\
%  &&\phantom{ -\Omega_{R}^{2}\int_{0}^{\infty}d\tau} \left(e^{\mathcal{L}_{0}(\tau)}\mathcal{\mathcal{P}}\rho(t-\tau)\right)a(-\tau)a^{\dagger}.\nonumber
% \end{eqnarray}
\begin{eqnarray}
\partial_{t}\mathcal{P}\rho & = & -i\left[H_{ph},\mathcal{P}\rho\right]+\mathcal{L}_{\Gamma}(\mathcal{P}\rho)\\
 &  & +\Omega_{R}^{2}\int_{0}^{\infty}d\tau\, e^{(i\omega_{at}-\Gamma_{pump}/2)\tau} a^{\dagger}\left(e^{\mathcal{L}_{0}(\tau)}\mathcal{\mathcal{P}}\rho(t-\tau)\right)a(-\tau)\nonumber\\
 &  & +\Omega_{R}^{2}\int_{0}^{\infty}d\tau\, e^{-(i\omega_{at}+\Gamma_{pump}/2)\tau} a^{\dagger}(-\tau)\left(e^{\mathcal{L}_{0}(\tau)}\mathcal{\mathcal{P}}\rho(t-\tau)\right)a\nonumber\\
 &  & -\Omega_{R}^{2}\int_{0}^{\infty}d\tau\, e^{(i\omega_{at}-\Gamma_{pump}/2)\tau}aa^{\dagger}(-\tau)\left(e^{\mathcal{L}_{0}(\tau)}\mathcal{\mathcal{P}}\rho(t-\tau)\right)\nonumber\\
 &  & -\Omega_{R}^{2}\int_{0}^{\infty}d\tau\, e^{-(i\omega_{at}+\Gamma_{pump}/2)\tau}\left(e^{\mathcal{L}_{0}(\tau)}\mathcal{\mathcal{P}}\rho(t-\tau)\right)a(-\tau)a^{\dagger}.\nonumber
\end{eqnarray}
At lowest order in $\Omega_{R}$, we can assume the interaction picture
density matrix in the convolution product to be constant, $\hat{\rho}(t-\tau)\simeq \hat{\rho}(t)$, i.e. $e^{\mathcal{L}_{0}\tau}\rho(t-\tau)\simeq \rho(t)$.
Making the trace over the bath we get :
\begin{eqnarray}
\partial_{t}\rho_{ph} & = & -i\left[H_{ph},\rho_{ph}\right]+\mathcal{L}_{\Gamma}(\rho_{ph})\\
 &  & +\Omega_{R}^{2}\int_{0}^{\infty}d\tau\, e^{(i\omega_{at}-\Gamma_{pump}/2)\tau}a^{\dagger}(-\tau)\rho_{ph}(t)a\nonumber\\
 &  & +\Omega_{R}^{2}\int_{0}^{\infty}d\tau\, e^{-(i\omega_{at}+\Gamma_{pump}/2)\tau}a^{\dagger}\rho_{ph}(t)a(-\tau)\nonumber\\
 &  & -\Omega_{R}^{2}\int_{0}^{\infty}d\tau\, e^{(i\omega_{at}-\Gamma_{pump}/2)\tau}aa^{\dagger}(-\tau)\rho_{ph}(t)\nonumber\\
 &  & -\Omega_{R}^{2}\int_{0}^{\infty}d\tau\, e^{-(i\omega_{at}+\Gamma_{pump}/2)\tau}\rho_{ph}(t)a(-\tau)a^{\dagger},\nonumber
\end{eqnarray}
then we can perform completely the integral and we get our final form
for the non Markovian master equation, which is local in time~: 
\begin{equation}
\partial_{t}\rho =  -i\left[H_{ph},\rho_{ph}\right]+\frac{\Gamma_{loss}}{2}\left[2a\rho a^{\dagger}-a^{\dagger}a\rho-\rho a^{\dagger}a\right] +\frac{2\Omega_{R}^{2}}{\Gamma_{pump}}\left[\tilde{a}^{\dagger}\rho a+a^{\dagger}\rho\tilde{a}-a\tilde{a}^{\dagger}\rho-\rho\tilde{a}a^{\dagger}\right],
\end{equation}
with 
\begin{equation}
\left\{ \begin{array}{l}
\tilde{a}=\frac{\Gamma_{pump}}{2}\int_{0}^{\infty}d\tau\, e^{(-i\omega_{at}-\Gamma_{pump}/2)\tau}a(-\tau) ,
\\
\tilde{a}^{\dagger}=\frac{\Gamma_{pump}}{2}\int_{0}^{\infty}d\tau\, e^{(i\omega_{at}-\Gamma_{pump}/2)\tau}a^{\dagger} (-\tau)=\left[\tilde{a}\right]^{\dagger} ,
\end{array}\right.
\end{equation}
where $ a(-\tau)$ means the photonic annihilation operator in the photonic hamiltonian interaction picture.

If $\ket f$ and $\ket f'$ are two eigenstates of the photonic hamiltonian with a photon number difference of one, we see that the matrix elements of the modified annihilation and creation operators $\tilde{a}$ and $\tilde{a}^{\dagger}$ involved in the emission process are :
\begin{equation}
\left\{ \begin{array}{l}
\bra f\tilde{a}^{\dagger}\ket{f'}=\frac{\Gamma_{pump}/2}{-i(\omega_{at}-\omega_{ff'})+\Gamma_{pump}/2}\bra fa^{\dagger}\ket{f'}\\
\bra{f'}\tilde{a}\ket f=\frac{\Gamma_{pump}/2}{i(\omega_{at}-\omega_{ff'})+\Gamma_{pump}/2}\bra{f'}a\ket f.
\end{array}\right.
\end{equation}
The non-Markovianity comes from the energy-dependence of the prefactors.

For several cavities the reasoning is exactly the same and we get the
multicavity master equation~:
% \begin{equation}
% \begin{array}{lll}
% \partial_{t}\rho & = & -i\left[H_{ph},\rho_{ph}\right]+\frac{\Gamma_{loss}}{2}\sum_{i=1}^{k}\left[2a_{i}\rho a_{i}^{\dagger}-a_{i}^{\dagger}a_{i}\rho-\rho a_{i}^{\dagger}a_{i}\right]\\
%  &  & +\frac{\Omega_{R}^{2}}{\Gamma_{pump}/2}\sum_{i=1}^{k}\left[\tilde{a}_{i}^{\dagger}\rho a_{i}+a_{i}^{\dagger}\rho\tilde{a_{i}}-a_{i}\tilde{a}_{i}^{\dagger}\rho-\rho\tilde{a}_{i}a_{i}^{\dagger}\right],
% \end{array}
% \end{equation}
\begin{equation}
\partial_{t}\rho = -i\left[H_{ph},\rho_{ph}\right]+\frac{\Gamma_{loss}}{2}\sum_{i=1}^{k}\left[2a_{i}\rho a_{i}^{\dagger}-a_{i}^{\dagger}a_{i}\rho-\rho a_{i}^{\dagger}a_{i}\right]+\frac{2\Omega_{R}^{2}}{\Gamma_{pump}}\sum_{i=1}^{k}\left[\tilde{a}_{i}^{\dagger}\rho a_{i}+a_{i}^{\dagger}\rho\tilde{a_{i}}-a_{i}\tilde{a}_{i}^{\dagger}\rho-\rho\tilde{a}_{i}a_{i}^{\dagger}\right],
\end{equation}
with 
\begin{eqnarray}
\bra f\tilde{a}_{i}\ket{f'} & = & \frac{\Gamma_{pump}/2}{i(\omega_{at}-\omega_{f'f})+\Gamma_{pump}/2}\bra fa_{i}\ket{f'} \label{eq:faifp} \\
\bra{f'}\tilde{a}_{i}^{\dagger}\ket f & = & \frac{\Gamma_{pump}/2}{-i(\omega_{at}-\omega_{f'f})+\Gamma_{pump}/2}\bra{f'}a_{i}^{\dagger}\ket f,\label{eq:fpaif}
\end{eqnarray}
where here also $\ket f$ and $\ket f'$ are two eigenstates of the many cavity photonic hamiltonian: once again the emission depends on the many body photonic dynamics via the prefactors in (\ref{eq:faifp}-\ref{eq:fpaif}).

\section{Lindblad form for the photonic master equation in the secular approximation\label{app:Lindblad-form}}

In this Appendix, we present the derivation of the Lindblad form Eq.~\ref{eq:photon_only_MCWF} for the photonic master equation including non markovian effects under the secular approximation. To do this, we calculate the matrix elements $\mathcal{L}_{em,\, f',\tilde{f}',f,\tilde{f}}$ of the emission superoperator coupling the term of the density matrix  in the eigenstate basis $\bra{f}\rho\tilde{\ket{f}}$ to $\bra{f'}\rho\tilde{\ket{f'}}$, under the assumption $\Delta\omega=\omega_{f',\tilde{f}'}-\omega_{f,\tilde{f}}\simeq 0$, as explained in Sec.~\ref{sec:photonic-lindblad-form}:
\subsection*{Calculation of the $\tilde{a}_i^\dagger\rho a_i+a_i^\dagger\rho \tilde{a}_i$ contribution:}
\begin{multline}
\bra{f'}\tilde{a}_i^\dagger\ket{f}\bra{f}\rho\tilde{\ket{f}}\tilde{\bra{f}} a_i \tilde{\ket{f'}}+\bra{f'}a_i^\dagger\ket{f}\bra{f}\rho\tilde{\ket{f}}\tilde{\bra{f}} \tilde{a}_i \tilde{\ket{f'}}\\
=\bra{f'}a_i^\dagger\ket{f}\bra{f}\rho\tilde{\ket{f}}\tilde{\bra{f}} a_i \tilde{\ket{f'}}
\left(\frac{\Gamma_{pump}/2}{-i(\omega_{at}-\omega_{f',f})+\Gamma_{pump}/2}
+\frac{\Gamma_{pump}/2}{i(\omega_{at}-\omega_{\tilde{f}',\tilde{f}})+\Gamma_{pump}/2}\right).
\end{multline}
Considering that under the approximation $\Delta\omega\simeq 0$, we have that $ \frac{\Gamma_{pump}/2}{-i(\omega_{at}-\omega_{f',f})+\Gamma_{pump}/2} \simeq \frac{\Gamma_{pump}/2}{-i(\omega_{at}-\omega_{\tilde{f}',\tilde{f}})+\Gamma_{pump}/2}$, we obtain thus the following contribution:
\begin{multline}
\bra{f'}\tilde{a}_i^\dagger\ket{f}\bra{f}\rho\tilde{\ket{f}}\tilde{\bra{f}} a_i \tilde{\ket{f'}}+\bra{f'}a_i^\dagger\ket{f}\bra{f}\rho\tilde{\ket{f}}\tilde{\bra{f}} \tilde{a}_i \tilde{\ket{f'}}\\
\simeq \bra{f'}a_i^\dagger\ket{f}\frac{\Gamma_{pump}/2}{\sqrt{(\omega_{at}-\omega_{f',f})^2+(\Gamma_{pump}/2)^2}}
\bra{f}\rho\tilde{\ket{f}}\frac{\Gamma_{pump}/2}{\sqrt{(\omega_{at}-\omega_{\tilde{f}',\tilde{f}})^2+(\Gamma_{pump}/2)^2}}\tilde{\bra{f}} a_i \tilde{\ket{f'}}\\
= \bra{f'}\bar{a}_i^\dagger\ket{f} \bra{f}\rho\tilde{\ket{f}}\tilde{\bra{f}} \bar{a}_i \tilde{\ket{f'}},
\end{multline}
with $\bar{a}_i$ defined in Eq.\ref{eq:crea_MCWF}. We see that the "imaginary" contribution cancels out, and that the "real" contribution has been divided in two multiplicative contributions on the left and the right of the density matrix.
\subsection*{Calculation of the $a_i\tilde{a}_i^\dagger\rho +\rho \tilde{a}_i a_i^\dagger$ contribution}
Let us calculate the left product:
\begin{multline}
\bra{f'}a_i\ket{f''}\bra{f''}\tilde{a}_i^\dagger\ket{f}\bra{f}\rho\tilde{\ket{f}}
=\bra{f'}a_i\ket{f''}\frac{\Gamma_{pump}/2}{-i(\omega_{at}-\omega_{f'',f})+\Gamma_{pump}/2}\bra{f''}a_i^\dagger\ket{f}\bra{f}\rho\tilde{\ket{f}}\\
=\bra{f'}a_i\ket{f''}\left[\frac{(\Gamma_{pump}/2)^2}{(\omega_{at}-\omega_{f'',f})^2+(\Gamma_{pump}/2)^2}-i\frac{(\omega_{f'',f}-\omega_{at})\Gamma_{pump}/2}{(\omega_{at}-\omega_{f'',f})^2+(\Gamma_{pump}/2)^2}\right]\bra{f''}a_i^\dagger\ket{f}\bra{f}\rho\tilde{\ket{f}}.
\end{multline}
Considering that under the approximation $\omega_{f',f}\simeq 0$, we have that  $\omega_{f'',f}\simeq \omega_{f'',f'}$, and so:
\begin{equation}
\frac{(\Gamma_{pump}/2)^2}{(\omega_{at}-\omega_{f'',f})^2+(\Gamma_{pump}/2)^2}\simeq \frac{(\Gamma_{pump}/2)^2}{\sqrt{(\omega_{at}-\omega_{f'',f})^2+(\Gamma_{pump}/2)^2}\sqrt{(\omega_{at}-\omega_{f'',f'})^2+(\Gamma_{pump}/2)^2}}.
\end{equation}
As a consequence:
\begin{multline}
\bra{f'}a_i\ket{f''}\bra{f''}\tilde{a}_i^\dagger\ket{f}\bra{f}\rho\tilde{\ket{f}}\\
\simeq -i \bra{f'}a_i\ket{f''}\frac{(\omega_{f'',f}-\omega_{at})\Gamma_{pump}/2}{(\omega_{at}-\omega_{f'',f})^2+(\Gamma_{pump}/2)^2}\bra{f''}a_i^\dagger\ket{f}\bra{f}\rho\tilde{\ket{f}}+\bra{f'}\bar{a}_i\ket{f''}\bra{f''}\bar{a}_i^\dagger\ket{f}\bra{f}\rho\tilde{\ket{f}} .
\end{multline}
Finally, let us calculate the right product:
\begin{multline}
\bra{f}\rho\tilde{\ket{f}}\tilde{\bra{f}}\tilde{a}_i\tilde{\ket{f''}}\tilde{\bra{f''}}a_i^\dagger\tilde{\ket{f'}}
=\bra{f}\rho\tilde{\ket{f}}\frac{\Gamma_{pump}/2}{i(\omega_{at}-\omega_{\tilde{f}'',\tilde{f}})+\Gamma_{pump}/2}\tilde{\bra{f}}\tilde{a}_i\tilde{\ket{f''}}\tilde{\bra{f''}}a_i^\dagger\tilde{\ket{f'}}\\
=\bra{f}\rho\tilde{\ket{f}}\left[\frac{(\Gamma_{pump}/2)^2}{(\omega_{at}-\omega_{\tilde{f}'',\tilde{f}})^2+(\Gamma_{pump}/2)^2}+i\frac{(\omega_{f'',f}-\omega_{at})\Gamma_{pump}/2}{(\omega_{at}-\omega_{\tilde{f}'',\tilde{f}})^2+(\Gamma_{pump}/2)^2}\right]\tilde{\bra{f}}\tilde{a}_i\tilde{\ket{f''}}\tilde{\bra{f''}}a_i^\dagger\tilde{\ket{f'}}
\end{multline}
As before, $\omega_{\tilde{f}'',\tilde{f}}\simeq \omega_{\tilde{f}'',\tilde{f}'}$, so:
\begin{eqnarray}
\frac{(\Gamma_{pump}/2)^2}{(\omega_{at}-\omega_{\tilde{f}'',\tilde{f}})^2+(\Gamma_{pump}/2)^2}&\simeq& \frac{(\Gamma_{pump}/2)^2}{\sqrt{(\omega_{at}-\omega_{\tilde{f}'',\tilde{f}})^2+(\Gamma_{pump}/2)^2}\sqrt{(\omega_{at}-\omega_{\tilde{f}'',\tilde{f}'})^2+(\Gamma_{pump}/2)^2}},\nonumber\\
\frac{(\omega_{\tilde{f}'',\tilde{f}}-\omega_{at})\Gamma_{pump}/2}{(\omega_{at}-\omega_{\tilde{f}'',\tilde{f}})^2+(\Gamma_{pump}/2)^2}&\simeq & \frac{(\omega_{\tilde{f}'',\tilde{f}'}-\omega_{at})\Gamma_{pump}/2}{(\omega_{at}-\omega_{\tilde{f}'',\tilde{f}'})^2+(\Gamma_{pump}/2)^2},
\end{eqnarray}
and thus
\begin{multline}
\bra{f}\rho\tilde{\ket{f}}\tilde{\bra{f}}\tilde{a}_i\tilde{\ket{f''}}\tilde{\bra{f''}}a_i^\dagger\tilde{\ket{f'}}
\simeq\bra{f}\rho\tilde{\ket{f}}\tilde{\bra{f}}\bar{a}_i\tilde{\ket{f''}}\tilde{\bra{f''}}\bar{a}_i^\dagger\tilde{\ket{f'}}\\
+i \bra{f}\rho\tilde{\ket{f}}\tilde{\bra{f}}a_i\tilde{\ket{f''}}\frac{(\omega_{\tilde{f}'',\tilde{f}'}-\omega_{at})\Gamma_{pump}/2}{(\omega_{at}-\omega_{\tilde{f}'',\tilde{f}'})^2+(\Gamma_{pump}/2)^2}\tilde{\bra{f''}}a_i^\dagger\tilde{\ket{f'}}.
\end{multline}
Here again, the real part has been divided in two multiplicative contributions, and the imaginary part has been swiped to the creation operator on the right. So wether we consider the contribution acting on the left or on the right of the density matrix, the imaginary contribution is always carried by the creation operator, so the density matrix is multiplied by the same operator on the right and the left up to a minus sign, which gives an anticommutator and thus an hamiltonian contribution due to the Lamb shift. 
\subsection*{Sum of the various contributions}
To summarize, keeping only relevant transitions we can consider that the emission dynamics is equivalent to a contribution $-i\left[\sum_{i} H_{lamb,i},\rho_{ph}\right]+ \bar{\mathcal{L}}_{em}$ in the master equation, with
\begin{equation}
\bar{\mathcal{L}}_{em}= \frac{\Gamma_{em}}{2}\sum_{i=1}^{k}\left[2\bar{a}_{i}^{\dagger}\rho_{ph} \bar{a}_{i}-\bar{a}_{i}\bar{a}_{i}^{\dagger}\rho_{ph}-\rho_{ph}\bar{a}_{i}\bar{a}_{i}^{\dagger}\right],
\end{equation}
\begin{equation}
 \bra{f'}\bar{a}_{i}^{\dagger}\ket{f}=\frac{\Gamma_{pump}/2}{\sqrt{(\omega_{at}-\omega_{f',f})^2+\left(\Gamma_{pump}/2\right)^2}}\bra{f'}a_{i}^\dagger\ket{f},
\end{equation}
\begin{equation}
 \bra{f'}H_{lamb,i}\ket{f}=\sum_{f''} \bra{f'}a_{i}\ket{f''}\left(\frac{(\omega_{f,f''}-\omega_{at})\Gamma_{pump}/2}{(\omega_{at}-\omega_{f,f''})^2+(\Gamma_{pump}/2)^2}\right)\bra{f''}a_{i}^\dagger\ket{f},
\end{equation}
which demonstrates the statements of Sec.\ref{sec:photonic-lindblad-form}.
\section{Exact stationary solution for Markovian case\label{app:Exact-stationary-solution}}

In this Appendix, we present a proof of our statements in Sec.\ref{sec:Markov}. We are looking for the steady state for the Markovian quantum dynamical process~:
\begin{equation}
\partial_{t}\rho =  -i\left[H,\rho(t)\right]+\mathcal{L}_{loss} + \mathcal{L}_{em},
\label{eq:photon_only_xxx}
\end{equation}
with standard Lindblad operators~:
\begin{equation}
\mathcal{L}_{loss} = \frac{\Gamma_{loss}}{2}\sum_{i=1}^{k}\left[2a_{i}\rho a_{i}^{\dagger}-a_{i}^{\dagger}a_{i}\rho-\rho a_{i}^{\dagger}a_{i}\right],
\end{equation}
\begin{equation}
\mathcal{L}_{em} =\frac{\Gamma_{em}}{2}\sum_{i=1}^{k}\left[2a_{i}^{\dagger}\rho a_{i}-a_{i}a_{i}^{\dagger}\rho-\rho a_{i}a_{i}^{\dagger}\right].
\end{equation}
We want to demonstrate that the following density matrix is an exact steady state~:
\begin{equation}
\rho_{\infty}=\sum_{N}\pi_{N}\mathcal{I}_{N} ,
\end{equation}
with 
\begin{equation}
\pi_{N}=A\left(\frac{\Gamma_{em}}{\Gamma_{loss}}\right)^{N}.
\end{equation}

First, since the hamiltonian preserves the total photon number, and that the density matrix is equal to the identity on each sub-space with a defined photon number, we get that $\com H{\rho_{\infty}}=0$. Second, for the Lindblad operators the non-hermitian hamiltonian terms have a simple action on the density matrix~:
\begin{equation}
\rho_{\infty}\sum_{i}a_{i}^{\dagger}a_{i}=\rho_{\infty}\op N=\op N\rho_{\infty}=\sum_{i}a_{i}^{\dagger}a_{i}\rho_{\infty},
\end{equation}
\begin{equation}
\rho_{\infty}\sum_{i}\underbrace{a_{i}a_{i}^{\dagger}}_{=a_{i}^{\dagger}a_{i}+1}=\underbrace{\rho_{\infty}(\op N+k)}_{=(\op N+k)\rho_{\infty}}=\sum_{i}a_{i}a_{i}^{\dagger}\rho_{\infty},
\end{equation}
where $k$ is the number of cavities.
We are left with the special terms of the form $a^{\dagger}\rho a$ and $a \rho a^{\dagger}$, for which we find that:
\begin{eqnarray}
\sum_{i}a_{i}^{\dagger}\rho_{\infty}a_{i} & = & \sum_{i}\sum_{N}\sum_{
\begin{array}{c}
f,\,\tilde{f}\,(N)\\
f'\,\tilde{f}'\,(N-1)
\end{array}
}\tilde{\ket f}\bra f\cdot \tilde{\bra f}a_{i}^{\dagger}\tilde{\ket{f'}}\underbrace{\tilde{\bra{f'}}\rho_{\infty}\ket{f'}}_{\pi_{eq(N-1)}\delta\tilde{f'},f'}\bra{f'}a_{i}\ket f\nonumber\\
 & = & \sum_{i}\sum_{N}\sum_{
 \begin{array}{c}
f,\,\tilde{f}\,(N)\\
f'\,(N-1)
\end{array}
}\tilde{\ket f}\bra f\cdot \pi_{N-1}\tilde{\bra f}a_{i}^{\dagger}\ket{f'}\bra{f'}a_{i}\ket f\nonumber\\\
 & = & \sum_{N}\sum_{
 \begin{array}{c}
f,\,\tilde{f}\,(N)\end{array}
}\tilde{\ket f}\bra f\cdot \pi_{N-1}\underbrace{\tilde{\bra f}\sum_{i}a_{i}^{\dagger}a_{i}\ket f}_{=N_{f}\delta_{f,f'}}\nonumber\\\
 & = & \sum_{N}\sum_{f(N)}N\pi_{N-1}\ket f\bra f.
\end{eqnarray}
and
\begin{eqnarray}
\sum_{i}a_{i}\rho_{\infty}a_{i}^{\dagger} = \sum_{N}\sum_{f(N)} (N+1+k)\pi_{N+1} \ket f\bra f.
\end{eqnarray}
If we sum all contributions together, it is immediate to see that we get a total zero contribution~:
\begin{multline}
\mathcal{L}_{loss}(\rho_{\infty})+\mathcal{L}_{em}(\rho_{\infty})= \\ =\sum_{N}\sum_{f(N)}\ket f\bra f \left(\underbrace{N\Gamma_{em}\pi_{N-1}-N\Gamma_{loss}\pi_{N}}_{=0}+\underbrace{(N+k)\Gamma_{loss}\pi_{N+1}-(N+k)\Gamma_{em}\pi_{N}}_{=0}\right)=0,
\end{multline}
which proves our statement. 

\section{Perturbative corrections to the coherences in the weakly non Markovian regime \label{app:quadratic coherences}}
In this Appendix we show that the lowest-order correction to the coherences between eigenstates (null in the Grand Canonical ensemble of Sec.\ref{sub:GC}) are quadratic in the inverse pumping rate $\Gamma_{pump}^{-1}$ and not linear as a naive pertubative expansion would suggest. To this purpose, we calculate the first order contributions to the coherences of the operator $\delta\mathcal{M}$ [defined in eqs.(\ref{perturbation}) and (\ref{deltacreation})] applied to the grand canonic density matrix and show them to be 0.
Let us calculate first the contribution of the first two terms~:
\begin{equation}
\sum_{i}\bra{f}\delta a_{i}^{\dagger}\rho_{\infty}a_{i}\ket{f'}=\sum_{i,\tilde{f},\tilde{f'}}\bra{f}\delta a_{i}^{\dagger}\tilde{\ket{f}}\tilde{\bra{f'}}a_{i}\ket{f'}\; \times\underbrace{\tilde{\bra{f}}\rho_{\infty}\tilde{\ket{f'}}}_{=\tilde{\bra{f}}\rho_{\infty}\tilde{\ket{f}}\delta_{\tilde{f},\tilde{f}'}}=\sum_{i,\tilde{f}}\bra{f}\delta a_{i}^{\dagger}\tilde{\ket{f}}\tilde{\bra{f}}a_{i}\ket{f'}\tilde{\bra{f}}\rho_{\infty}\tilde{\ket{f}} \\
\end{equation}
In the same way~:
\begin{equation}
\sum_{i}\bra{f}a_{i}^{\dagger}\rho_{\infty}\delta a_{i}\ket{f'} =\sum_{i,\tilde{f}}\bra{f}a_{i}^{\dagger}\tilde{\ket{f}}\tilde{\bra{f}}\delta a_{i}\ket{f'}\tilde{\bra{f}}\rho_{\infty}\tilde{\ket{f}}
\end{equation}
Then we know that 
\begin{equation}
\bra{f}\delta a_{i}^{\dagger}\tilde{\ket{f}}=-\frac{i(\omega_{f\tilde{f}}-\omega_{at})}{\Gamma_{pump}}\bra{f}a_{i}^{\dagger}\tilde{\ket{f}}+\mathcal{O}\left(\frac{1}{\Gamma_{pump}}\right)^{2}.
\end{equation}
Let us choose a reference state $\ket{f_0}$ with the same photon number as $\tilde{\ket{f}}$. Then $\tilde{\bra{f}}\rho_{\infty}\tilde{\ket{f}}=\bra{f_0}\rho_{\infty}\ket{f_0}+\mathcal{O}(\Gamma_{pump}^{-1})$. All these additional terms give second order contributions, and we do not consider them.
Thus to the first order~:
\begin{equation}
\begin{array}{l}
\sum_{i}\bra{f}\delta a_{i}^{\dagger}\rho_{\infty}a_{i}+a_{i}^{\dagger}\rho_{\infty}\delta a_{i}\ket{f'}\\
\phantom{\sum_{i}\bra{f}\delta a_{i}^{\dagger}\rho_{\infty}}=\sum_{i,\tilde{f}}\bra{f}a_{i}^{\dagger}\tilde{\ket{f}}\tilde{\bra{f}}a_{i}\ket{f'}\bra{f_0}\rho_{\infty}\ket{f_0}\\
\phantom{\sum_{i}\bra{f}\delta a_{i}^{\dagger}\rho_{\infty}=\sum_{i,\tilde{f}}}\frac{-i(\omega_{f\tilde{f}}-\omega_{f'\tilde{f}})}{\Gamma_{pump}}\\
\phantom{\sum_{i}\bra{f}\delta a_{i}^{\dagger}\rho_{\infty}}=A\frac{-i\omega_{ff'}}{\Gamma_{pump}}\sum_{i,\tilde{f}}\bra{f}a_{i}^{\dagger}\tilde{\ket{f}}\tilde{\bra{f}}a_{i}\ket{f'}\\
\phantom{\sum_{i}\bra{f}\delta a_{i}^{\dagger}\rho_{\infty}}=A\frac{-i\omega_{ff'}}{\Gamma_{pump}}\sum_{i}\bra{f}a_{i}^{\dagger}a_{i}\ket{f'}\\
\phantom{\sum_{i}\bra{f}\delta a_{i}^{\dagger}\rho_{\infty}}=A\frac{-i\omega_{ff'}}{\Gamma_{pump}}\underbrace{\bra{f}N\ket{f'}}_{=N_{f}\delta_{ff'}}\\
\phantom{\sum_{i}\bra{f}\delta a_{i}^{\dagger}\rho_{\infty}}=A\frac{-i\omega_{ff'}}{\Gamma_{pump}}N_{f}\delta_{ff'}\\
\phantom{\sum_{i}\bra{f}\delta a_{i}^{\dagger}\rho_{\infty}}=0.
\end{array}
\end{equation}
A similar reasoning allows to show that
\begin{equation}
\sum_{i}\bra{f}a_{i}\delta a_{i}^{\dagger}\rho_{\infty}+\rho_{\infty}\delta a_{i}a_{i}^{\dagger}\ket{f'}=0
\end{equation}
which completes our proof.

\section{Further numerical validation of the photonic master equation \label{app:numerical_validation}}
\label{app:valid}

\subsection*{One cavity case}

\begin{figure}
\begin{center}
\begin{tabular}{c}
\includegraphics[width=0.32\columnwidth,clip]{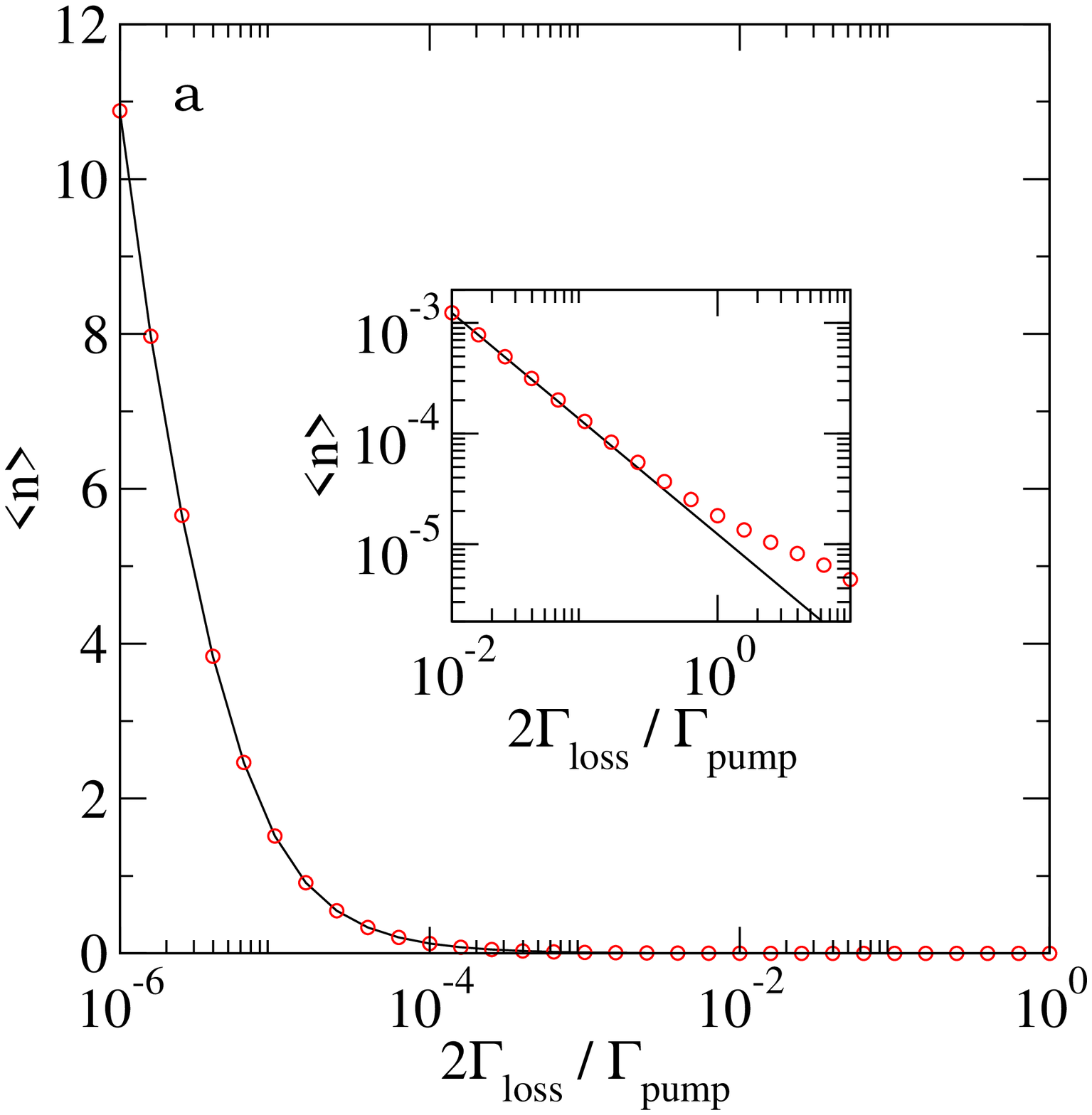}
\hspace*{0.1\columnwidth}
\includegraphics[width=0.3\columnwidth,clip]{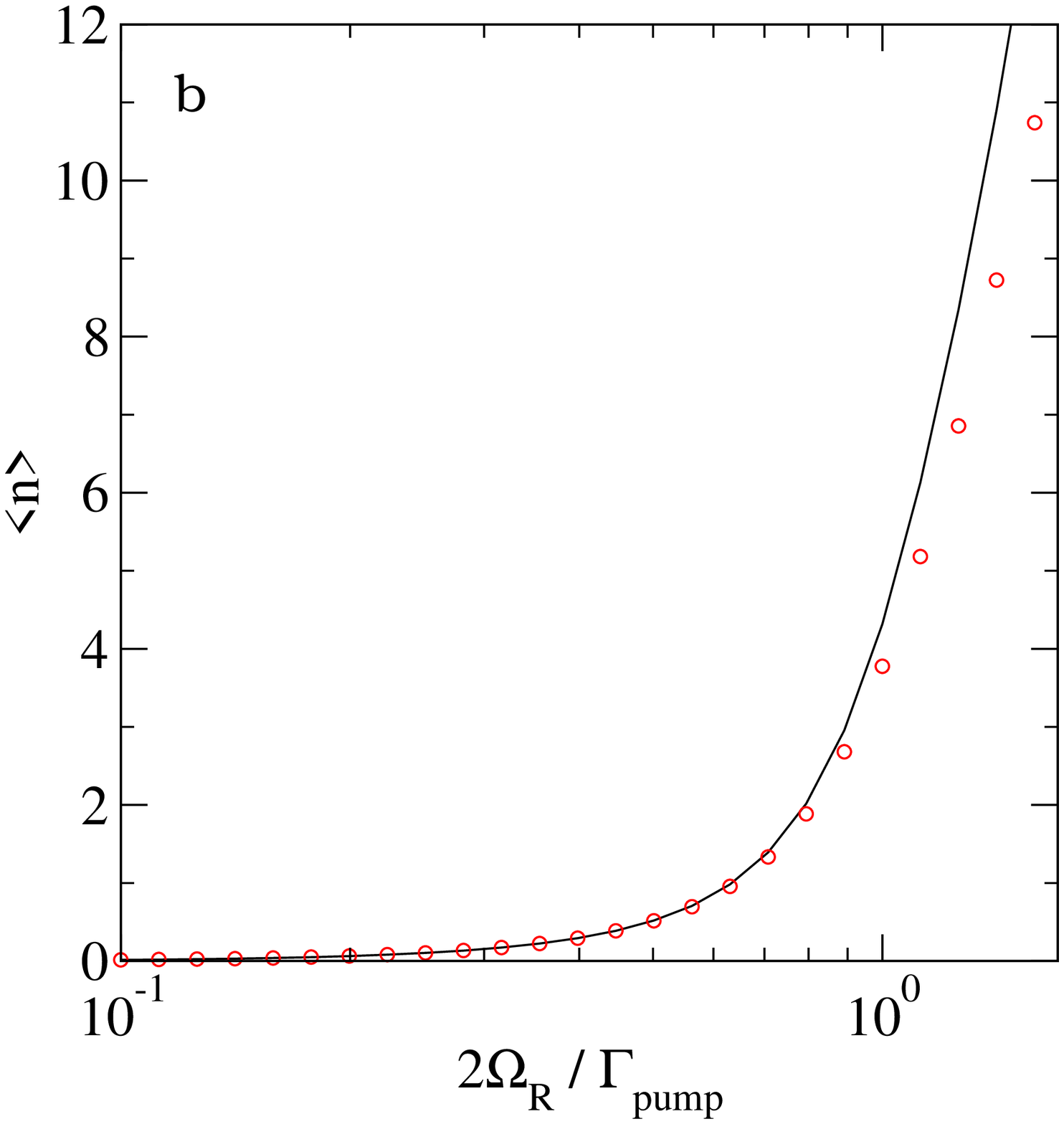}
\end{tabular}
\end{center}
\caption{Comparison of the analytical prediction of the photonic theory (solid black line) to the numerical solution of the full atom-cavity master equation (open red points). Stationary value of the average number of photons as a function of the photon loss rate $\Gamma_{loss}$ ({left}) and of the atom-cavity coupling $\Omega_R$ ({right}). Parameters~: $2U/\Gamma_{pump}=2$, $2\Omega_{R}/\Gamma_{pump}=0.02$ [left panel (a)]; $2U/\Gamma_{pump}=0.6$, $2\Gamma_{loss}/\Gamma_{pump}=0.02$ [right panel (b)]. In all panels, $2\delta/\Gamma_{pump}=8$. 
\label{fig:theoryonetest}}\end{figure}
%a
%U=1;
%G_ex=0;
%P_ex=1;
%g=0.01;
%omega_ph=5;     
%omega_at=1;
%b
%U=0.3;
%G_ex=0;
%P_ex=1;
%G_loss=0.01;
%omega_ph=5;     
%omega_at=1;

% %verage number of photons in function of the atom-cavity coupling $\Omega_{R}$ in units of the pump rate $\Gamma_{pump}/2$, given
% by numerical simulation of the atom/cavity problem, and by the master
% equation for photon degrees of freedom given by projective method.
% For small coupling simulation and theory give identical results. When
% the coupling becomes comparable to the pumping, divergences start
% to appear.\label{fig:theoryonetest2}

Here we compare the analytical prediction for the stationary state of the atom-cavity system discussed in Sec.\ref{sec:onecavity} to a numerical solution of the full master equation Eq.(\ref{eq:evinitio}). For example, in the left panel of Fig.\ref{fig:theoryonetest} the stationary value for the average photon number is plotted as a function of the photon loss rate $\Gamma_{loss}$. As expected, the purely photonic approach based on the projective method gives very accurate results as long as the pump rate $\Gamma_{pump}$ (i.e. the inverse autocorrelation time of the atomic bath) is much faster than the loss rate $\Gamma_{loss}$. 

A similar plot of the average photon number as a function of the atom-cavity coupling $\Omega_{R}$ is shown in the {right} panel. Outside the small $\Omega_R$ regime, the photonic theory tends to overestimate the photon number. This deviation can be explained as the theory assumes the atoms to be always in their excited state ready for emission and neglects the possibility of an atom reabsorbing the emitted photon before being repumped to the excited state.

% Note that in both panels no constraint needs to be put on the parameters of the free photonic hamiltonian: the purely photonic theory well matches the numeric prediction also when the detuning $\delta=\omega_{cav}-\omega_{at}$ and/or the nonlinearity $U$ are comparable to the pumping rate $\Gamma_{pump}$. This fact will play a crucial role when trying to exploit the energy selectivity of the emission process to generate specific complex many-body states.

\subsection*{Two cavity case}

\begin{figure}
\begin{center}
\includegraphics[width=0.31\columnwidth,clip]{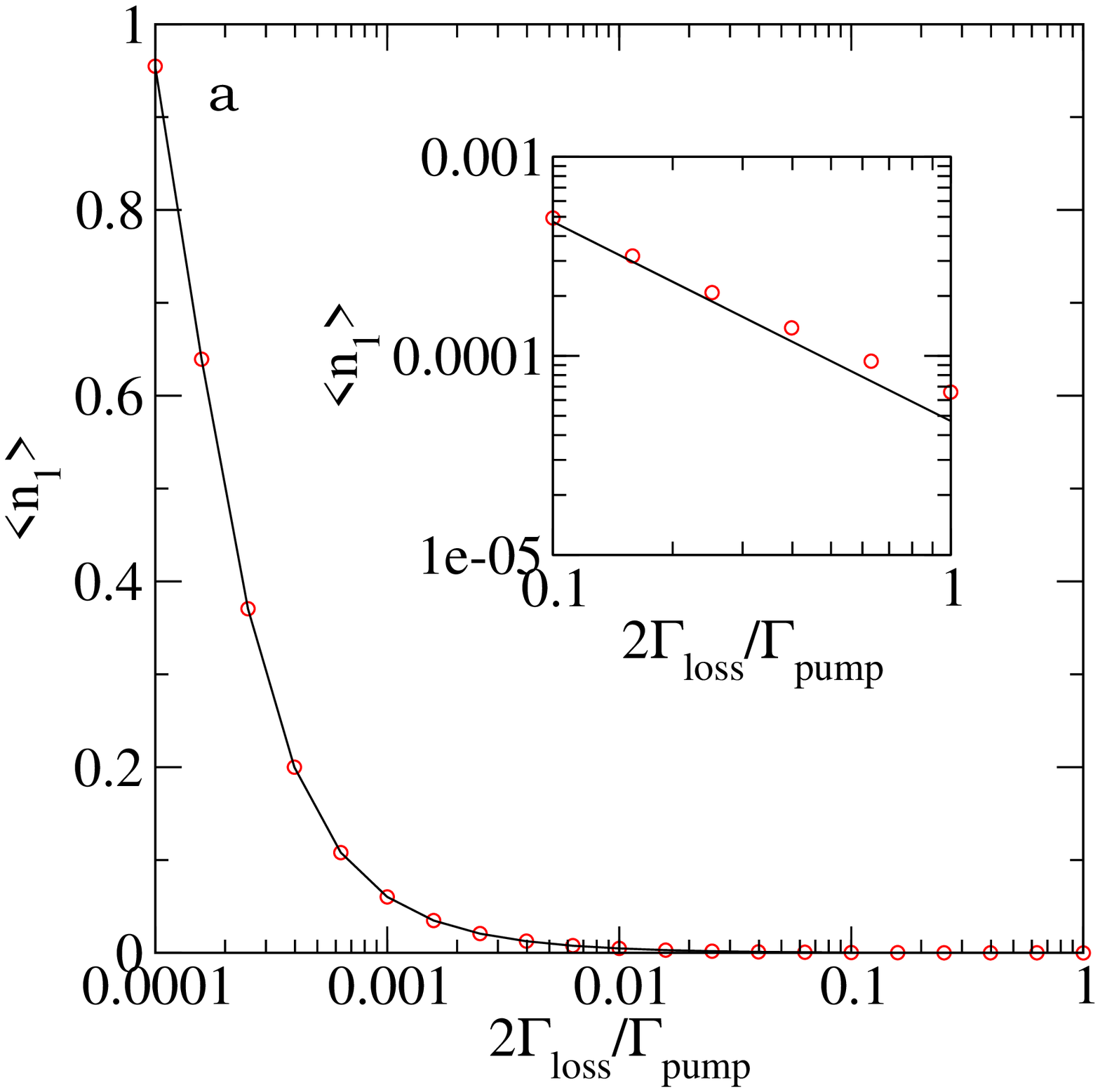}
\includegraphics[width=0.3\columnwidth,clip]{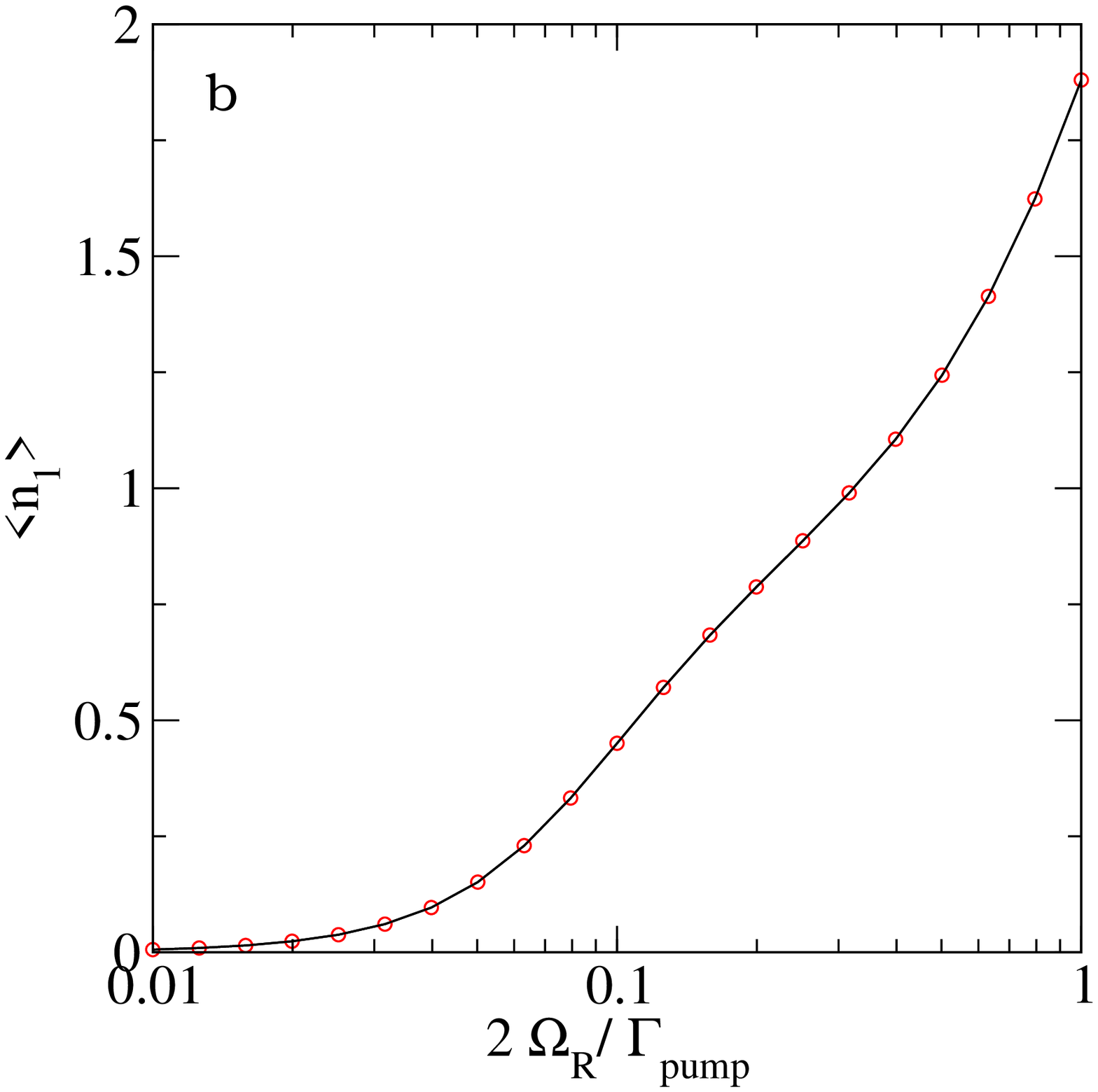}
\end{center}
\caption{Comparison of the analytical prediction of the photonic theory (solid black line) to the numerical solution of the full atom-cavity master equation (open red points) for a two-cavity system. Stationary value of the average number of photons in the first cavity as a function of the photon loss rate $\Gamma_{loss}$ ({left}) and of the atom-cavity coupling $\Omega_R$ ({right}). Parameters: $2U/\Gamma_{pump}=7$, $2\Omega_{R}/\Gamma_{pump}=0.02$, (left \textbf{a)} panel); $2U/\Gamma_{pump}=28$,  $2\Gamma_{loss}/\Gamma_{pump}=0.002$ (right \textbf{b)} panel). In all panels, {$2J/\Gamma_{pump}=4$} and $\delta=0$.
\label{fig:theorytwotest}}
\end{figure}
%a
%U=3.5;
%J=1;
%G_ex=0.00001;
%P_ex=1;
%G_ph=0.001;
%g=0.01;
%omega_ph=1;      % cavity frequency
%omega_at=1;
%
%b
%U=14;
%J=1;
%G_ex=0.00001;
%P_ex=1;
%G_ph=0.001;
%g=0.0001;
%omega_ph=1;      % cavity frequency
%omega_at=1;

Here we give further validation to the purely photonic description used in Sec.\ref{sec:Arrays-of-cavities} by comparing its predictions with the numerical results for the full atom-cavity master equation in a two cavity case. An example is shown in Fig.\ref{fig:theorytwotest}: as in the single cavity case, the agreement is excellent at large $\Gamma_{pump}$ and gets deteriorated when $\Gamma_{pump}$ is decreased to values comparable to $\Gamma_{loss}$ [panel (a)]. The situation is even more favourable in panel (b), where the deviations that are expected for larger $\Omega_R$ are suppressed by the strong nonlinearity. These numerical results offer a further validation of the analytical approximations underlying our the photonic approach.

%\bibliography{mybibfile}

\end{document}